\documentclass[twocolumn,tighten]{aastex63}
\usepackage{amssymb, amsmath,framed}

\usepackage{bm} 
\usepackage{multirow}
\usepackage{gensymb}
\usepackage{physics}
\expandafter\ifx\csname package@font\endcsname\relax\else
 \expandafter\expandafter
 \expandafter\usepackage
 \expandafter\expandafter
 \expandafter{\csname package@font\endcsname}
\fi
\def\bq{\begin{equation}}
\def\eq{\end{equation}}
\def\bqy{\begin{eqnarray}}
\def\eqy{\end{eqnarray}}
\def\p{\partial}
\def\ph{\phi}

\def\p{\partial}

\def\R{\mathbb{R}}

 \def\q#1#2{q^{\hspace{1pt}#1}_{\,,\hspace{1pt}#2}}
 \def\a#1#2{a^{\hspace{1pt}#1}_{\,,\hspace{1pt}#2}}
 \def\d#1#2{\delta^{\hspace{1pt}#1}_{\,,\hspace{1pt}#2}}

\begin{document}
\title{\large{Science from the \textit{In Situ} Exploration of the Proxima Centauri System}}

\correspondingauthor{T. Marshall Eubanks}
\email{tme@space-initiatives.com}

\author[0000-0001-9543-0414]{T. Marshall Eubanks}
\affiliation{Space Initiatives Inc, Princeton, WV 24740, USA}  

\author[0000-0002-3136-6537]{Jean Schneider}
\affiliation{LUX, Paris Observatory, Meudon 92190 France.}
 
\author[0000-0002-1978-8243]{Bruce Bills}
\affiliation{Jet Propulsion Laboratory, California Institute of Technology, Pasadena, CA, USA}

\author{W. Paul Blase}
\affiliation{Space Initiatives Inc, Princeton, WV 24740, USA}

\author[0000-0003-1763-6892]{Andreas M. Hein}
\affiliation{Initiative for Interstellar Studies, London, UK}
\affiliation{SnT, University of Luxembourg, Luxembourg}

\author[0000-0003-0626-1749]{Pierre Kervella}
\affiliation{LIRA, Paris Observatory, Meudon 92190 France}

\author[0000-0003-1116-576X]{Adam Hibberd}
\affiliation{Institute for Interstellar Studies (i4is US), Oak Ridge, TN, USA}

\author[0000-0003-3924-7935]{Robert G. Kennedy III}
\affiliation{Institute for Interstellar Studies (i4is US), Oak Ridge, TN, USA}

\author[0000-0002-2685-9417]{Manasvi Lingam}
\affiliation{Department of Aerospace, Physics and Space Sciences, Florida Institute of Technology, Melbourne, FL 32901, USA}
%\affiliation{Department of Chemistry and Chemical Engineering, Florida Institute of Technology, Melbourne, FL 32901, USA}
\affiliation{Department of Physics and Institute for Fusion Studies, The University of Texas at Austin, Austin, TX 78712, USA}

\author[0000-0002-4973-8896] {Philip Lubin}
\affiliation{University of California at Santa Barbara, Santa Barbara, CA, USA}

\author[0000-0001-6397-5516]{Philip D. Mauskopf}
\affiliation{Department of Physics and School Of Earth and Space Exploration, Arizona State University, Tempe, AZ 85287, USA}

\author[0000-0003-4689-4997]{Thomas J. Mozdzen}
\affiliation{School of Earth and Space Exploration, Arizona State University, Tempe, Arizona 85287, USA.}

\author[0009-0005-0203-9229]{Richard M. Scott}
\affiliation{School of Earth and Space Exploration, Arizona State University, Tempe, Arizona 85287, USA.}

\author[0000-0003-4255-9497]{Slava G. Turyshev}
\affiliation{Jet Propulsion Laboratory, California Institute of Technology, Pasadena, CA, USA}

\begin{abstract} 
In the future interstellar exploration at near-relativistic speeds will be possible using beamed energy laser propulsion. With this, spacecraft as small as gm mass picospacecraft  become candidates for the exploration of deep space, with a trade space of velocity and mission duration versus mass. 
Here, we  examine the potential science return from  interstellar expeditions with Coracle\texttrademark  laser-sail picospacecraft swarms and show how even with fast flybys at near
relativistic velocities, a picospacecraft swarm could deliver gigapixel resolution of the target exoplanets. Our mission target is the planet Proxima  b in the habitable zone (HZ) of the red dwarf Proxima Centauri, the 
tertiary (and nearest) component
of the nearest star system, $\alpha$ Centauri. We explore science returns from such an expedition, both \textit{en route} to Proxima and  at the Proxima system, and  conclude that  initial small spacecraft expeditions  would provide a substantial science return, including the ability to detect surface biology or a technological civilization, should either or both  be established on the target planet.\\
%% leave the ending \\ in!
%%
\end{abstract}

\section{Introduction} 

 With a large scale concerted effort, it should be possible to send relativistic probes to nearby star systems in the next few decades using laser beamed propulsion \citep{Lubin-2016-a,Parkin-2018-a,Lin-et-al-2025-a}. Given the predicted rate of development of laser-sail technology the initial relativistic probes will likely be picospacecraft with a mass of a few grams each \citep{Crawford-2018-a}. Obtaining and returning an adequate data return from such probes can be advanced by launching many probes in succession and having these cooperate together in a swarm. A given laser launch system can launch a wide range of masses, with the choice of spacecraft mass being determined by the mission requirements, including time to target,  as well as by the maturation of low-mass spacecraft technologies. Since the laser array is modular and scalable, options ranging from one or more extremely low mass (gram-scale) to fairly large (many kilogram-scale) spacecraft become feasible once the launch system is available.   
 
 The scaling of boost velocity with spacecraft mass is discussed in detail by \cite{Lubin-2016-a}, who finds that the probe velocity, V, scales as V $\propto$ mass$^{-1/4}$ for a given reflector material. A laser system that could  accelerate 3.6 gm probes to 0.2 c could thus accelerate 100 gm probes to 0.087 c, 1 kg probes to 0.049 c or 100 kg probes to 0.015 c. Probes in the 1 - 100 kg mass range, accelerated to velocities of a thousand au / year or more, could explore almost every scientific target in the solar system, including the solar gravitational lens (SGL) focal region starting at 550 au \citep{Turyshev-et-al-2020-b}, and the nearby Oort cloud beyond.
 
There is a trade-off between sending many low mass probes versus fewer larger mass probes and this trade space is complicated by a number of issues including the ability to communicate over the vast distances required. While larger mass missions are slower for the same laser infrastructure, they are generally more capable in terms of instrumentation, power and communication. In general, we envision the same laser launch facility will allow a wide range of possible missions, such as slower missions with higher probe masses to solar system targets,  single probe scouting missions into interstellar space, and large swarm missions to distant targets, including to the SGL focal lines \citep{Turyshev-et-al-2020-b}, to free floating nomadic planets near the solar system \citep{Lingam-et-al-2023-b}, to material captured in the Sun-Galactic Lagrange points 3.8 light  years  away \citep{Belbruno-Green-2024-a}, and to the stellar systems nearest to the Sun \citep{Heller-et-al-2017-a}. 

\section{A Mission to Proxima Centauri}
\label{sec:mission-to-Proxima}

As a test case this paper focuses on science goals for a picospacecraft swarm mission to the Proxima Centauri (or Proxima) system, the nearest star to the Sun at a distance of 4.24 light years (ly), primarily to explore its planet Proxima b \citep{Eubanks-et-al-2024-b}.
Tables \ref{table:Centauri_System} and \ref{table:Centauri_Planets} provide the basic parameters of this system.
Proxima is an M5.5 dwarf and flare star with a radius of $\sim$98,100 km 
Proxima~b is a rocky planet in Proxima's HZ at a semi major axis (a) f 0.049 au; its assumed radius is 7600 km. Proxima~d is a rocky planet at a = 0.029 au, at the edge (at best) of Proxima's HZ but likely too close to Proxima to be habitable; its assumed radius is 5200 km.	Proxima~c is still unconfirmed; if it exists it  would be  well outside the Proxima HZ.

\subsection{Spacecraft Swarm Missions to Proxima Centauri}
\label{subsec:swarm-missions}

The mission would be conducted by an operationally coherent swarm
of 1000 ``Coracle'' light sails  launched over no more than 1 month
at a planned  velocity of 0.2 c, which  yields an apparent (Earth proper time) mission duration of  21.2 years. These probes will be assembled into a coherent swarm using drag from the Interstellar Medium (ISM) ``wind'' (the ISM impacting each probe at 0.2 c) to adjust the various probe trajectories in order to bring them into relatively close proximity (order 100,000 km transverse separation).
 The data return rate goal after 1 year of data return, 4.44 ly from Earth, is 0.9 kbps. This mission would return a total of $\sim$3.4 gigabytes of encounter data back to Earth.

\begin{figure}[ht]
\centering
\includegraphics[width=\columnwidth]{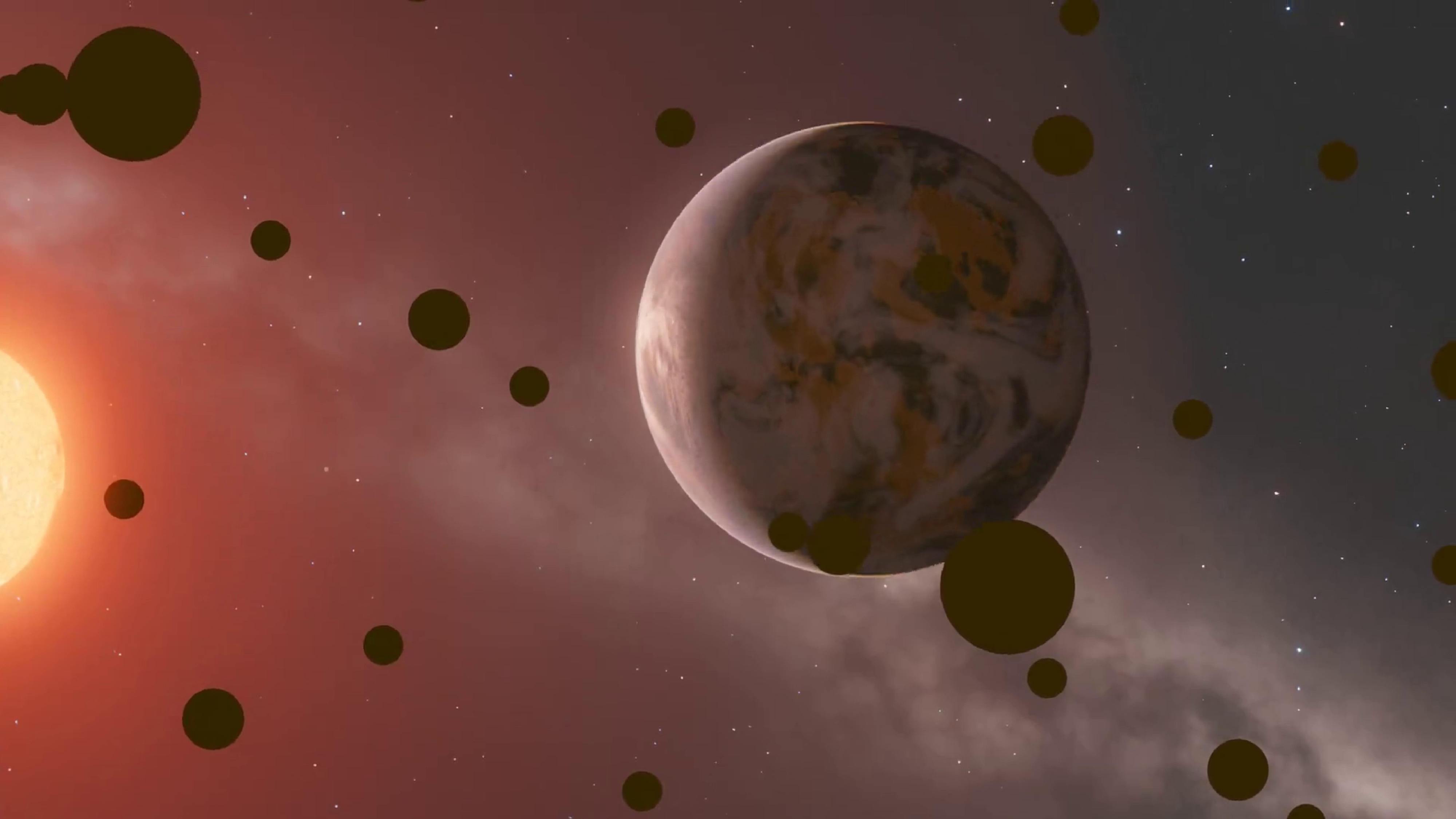}
%%\plotone{Moon_latitude_temp_11.eps}
%%
\caption{Artist's impression of the approach of a swarm towards Proxima b; at this point, a few seconds before closest approach, the swarm could be examining the planet's night-side for techno- or 
bioluminescence. (This image is based on the artistic work of Dr. Mark A. Garlick.)
\label{fig:Flyby-1}
}
\end{figure}

\begin{figure*}[ht]
\includegraphics[width=\textwidth]{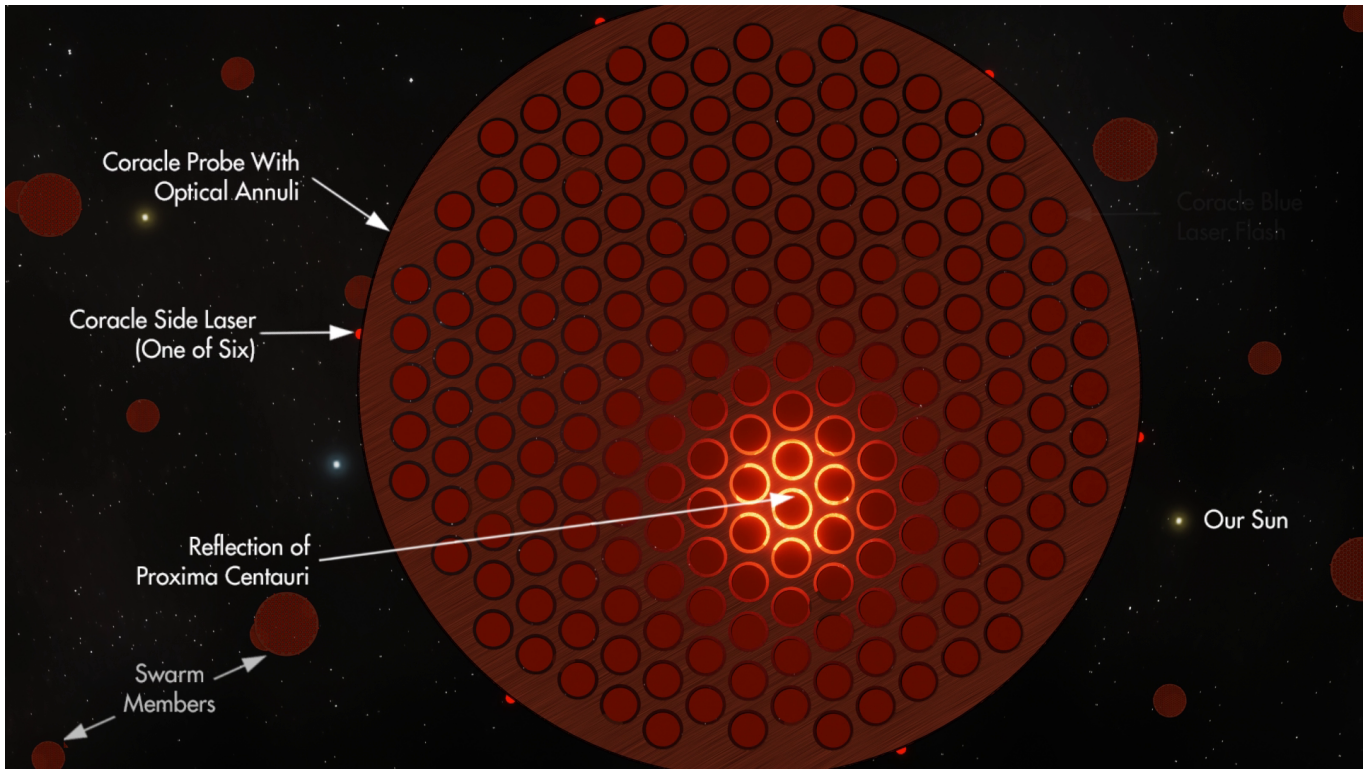}
\caption{Artist's impression of a Coracle approaching Proxima b (and reflecting the light of Proxima Centauri). The  12,000-nm intra-swarm ``Side Lasers'' (see Subsection \ref{subsec:Intra-swarm-comms}) are for intra-swarm probe-to-probe communications. Each round ring on the top (instrumentation) side of the sail visible here is the 200 mm  annulus aperture of a folded optic camera (see Figure \ref{fig:Optics_Cross_section_2} and discussion) shared between imaging and communications with Earth at 432/539-nm. Conceptual artwork by Mark Garlick. (Note: Seeing other probes apparently nearby at encounter is artistic license!)}
\label{fig:Flyby-2}
\end{figure*}

\begin{figure}[ht]
\centering
\includegraphics[width=\columnwidth]{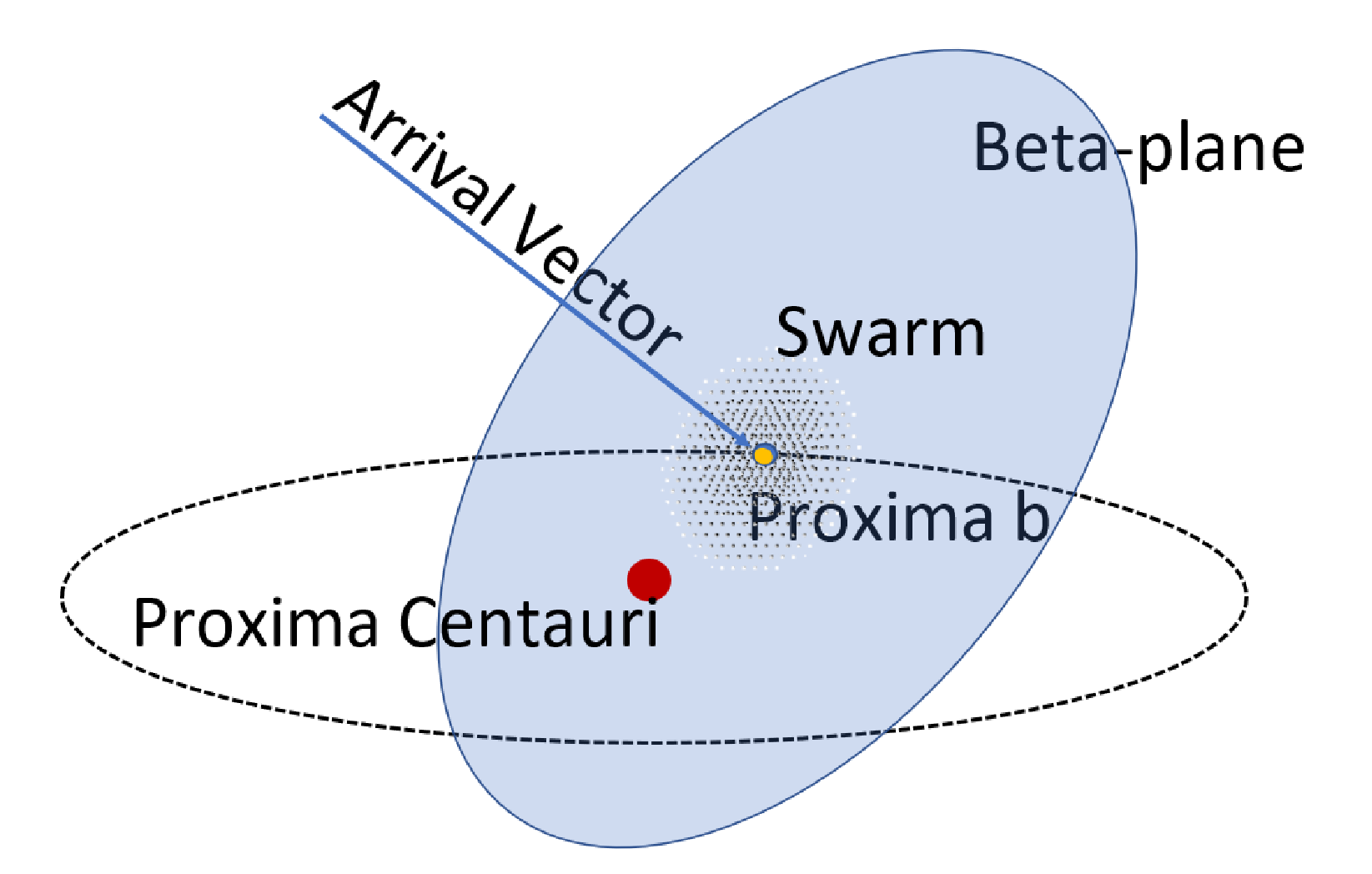}
%%\plotone{Moon_latitude_temp_11.eps}
%%
\caption{The beta-plane of a swarm flyby of Proxima Centauri b, with the swarm shown lying in that plane. (Note that the planned swarm dispersion is much smaller than is indicated in this artist's impression, and that in practice the swarm will not be exactly centered on Proxima b's position due to ephemeris errors.)
\label{fig:Flyby-3}
}
\end{figure}

Figure \ref{fig:Flyby-1} shows an artist's impression, from a NASA video\footnote{\url{https://www.youtube.com/watch?v=XXW_keR5OIM&t=2s}},  of the Coracle swarm approaching Proxima b.
Figure \ref{fig:Flyby-2} shows an annotated  artist's impression, from the same video, of a Coracle approaching Proxima. The swarm is distributed in the ``beta-plane'' (Figure \ref{fig:Flyby-3}), orthogonal to the direction of motion. 

An advantage of swarms with large numbers of probes is that they can provide a high probability that at least some  probes will pass close on both sides of the target. There are multiple ways to utilize the launch capability to send out a number of probes; while this paper focuses on the exploration of the planet Proxima b, there is of course other interesting science to be done in the Proxima system, and sub-swarms devoted to close passages of the other planets in the system, to the study of Proxima as a star, or to other goals, are likely to be a part of any initial exploration missions. Figure \ref{fig:Flyby-3} shows the beta-plane of a single swarm encounter for a  time coherent swarm. 

%%%
%%%%%
%%%%%

\subsection{Swarm Communications Over Interstellar Distances}
\label{subsec:swarm-comms}

In all cases enabling feasible communications back to Earth is critical and in this there is a trade space that must be carefully 
considered. Possibilities include
\begin{enumerate}
    \item \textbf{Independent probes} The probes are sent sequentially as a string  of isolated autonomous units transmitting data independently. This relaxes constraints on when the probes are sent, but will greatly limit any multi-probe studies, and may result in the transmission of less data, and also of redundant data back to Earth.  This is of course also the fall-back option if swarm coherence cannot for any reason  be obtained. 
    
    \item \textbf{Time coherent swarm} The probes act as a temporally coherent (but not optically phase coherent) swarm sending communications pulses which arrive at Earth simultaneously in a ``wall of light.'' 
    This requires that the swarm knows the relative position of each probe, that the swarm members have synchronized clocks, and that the location of the Earth relative to the Earth is known well enough that each communication packet of photons can be sent to arrive at Earth within a pulse  time window, allowing higher Signal to Noise Ratio (SNR) \citep{Eubanks-et-al-2023-b}. For the nominal nanosecond pulse widths, this requires nanosecond level positioning and timing of each individual probe.  It is the latter approach of temporal data transmission coordination between the probes that potentially allows a significant increase in data throughput.
    
    For PPM (Pulse Position Modulation) laser communications, the time gating of the photon packet arrival is critical to keep a relatively short photon arrival coordination. This is complicated by the generally low power of the probe transmitter, detector dark count, sky backgrounds, host star background, etc. \citep{Messerschmitt-2014-a}.   While none of this is simple to accomplish, it potentially offers a significant increase in effective data rate and graceful degradation in the case of individual probe failure. 
    
    \item \textbf{Sparse Phased Array} An even more exotic form of swarm coordination would be to form a true sparse phased array approach in which each probe is phase locked to all the others to form a synthetic sparse aperture array. The advantage of this latter approach is that the synthesized beam pattern in the main lobe at Earth is equivalent to that of the single transmission reflector with area equal to the sum of the areas of all the probes, although this would be a sparse array and the beam shape would not be the same as the beam formed by a solid antenna with the same extent. Note that this approach would require maintaining the positions of the probe members at the few 100 nm level or better, roughly 6 orders of magnitude better than time coherent swarm approach. We do not consider this last sparse phased array approach further in this paper due to the extreme difficulty of phase coordination across the swarm. 
\end{enumerate}
%%\section[The $\alpha$ Centauri System]{The Alpha Centauri System}

\section{Mission Concept of Operations}
\label{section:ConOps}

The concept of operations (Conops) for an interstellar swarm mission involves a set of  steps that have to occur in proper sequence. All times here are in Earth proper time; after the boost phase probe clocks will be kinematically redshifted, which amounts to an $\sim$5-month time shift over 20 years. Table \ref{table:swarm_parameters} provides details on the basic engineering parameters assumed for the swarm; we assume that launch losses, regular losses during the voyage, the sacrifice of probes for communications needs (or to observe directly forward \textit{en route}), and losses during the Proxima encounter itself, will amount to 70\% of the original swarm, leaving 300 probes for communications back to Earth.
\begin{itemize}
    \item
    Launch will occur in a ``face-up'' orientation, with the sail-side facing the laser and the instrument-side facing away from Earth, during a nominal powered flight duration of 8 minutes \citep{Parkin-2018-a}. 
    \item 
    As soon as the boost phase ends, each probe will rapidly conduct a 90$\degree$ rotation to travel edge-on, to minimize the erosion and collision risk from the interplanetary medium and dust (Subsection \ref{subsec:damage-reduction}), and will exit the solar system in that attitude. The direction to Proxima Centauri is 53.065$\degree$ from the nose direction of the heliopause \citep{Spitzer-et-al-2024-a}, which indicates that heliopause passages should occur at distances comparable to the Voyager spacecraft heliopause passages at 125.6 and 119 au.
    We therefore assume that the solar system risk zone extends out to 125 au, which would take $\sim$3.6 days to transit at 0.2 c. 
    \item 
    During this early period  positioning and ranging from Earth can be quite accurate; a milli-arc-second (mas) angular accuracy should be easily achievable at 125 au, yielding a transverse accuracy of $\sim$100 km, and very accurate two-way ranging would be possible.  
    The intra-swarm rim laser system would easily suffice for communications with Earth, and tracking done from Earth can be rapidly communicated to the swarm members.
    \item 
    Boosting the entire swarm into space could take between 6 days and  $\sim$1 month \citep{Eubanks-et-al-2023-a}; we use a default nominal mission boost phase duration of 6 days, leading to a maximum swarm radial dispersion of $\sim$208 au. A time on target technique will be used to coalesce the entire swarm in position by increasing the launch velocity of each new probe, so that the later launched probes catch up to the earlier ones. 
    With a 6 day boost duration, a $\Delta$v$_{s}$ spread of $\sim$352 km s$^{-1}$ will need to be imposed at launch between the beginning and the end of the deployed swarm.
%%     200 au / 10^8 seconds = 2.992 × 10^13 / 10^8 = 3 x 10^5 m/s
%%     3 x 10^5 / 4 x 10^-3 = 7.5 x 10^7 seconds = 865 days = 2.37 years
    \item 
    It is not possible to increase speeds with drag from the ISM wind caused by the probe's velocity; we use ISM drag to implement a velocity on target technique, slowing down the later launched probes so that velocities come to match as probes approach each other. 
    Once the solar system risk zone is passed this technique will be initiated by rotating swarm members into a ``face-down'' sail-side up configuration, increasing the drag by having the sail-side face into the ISM wind.  In the face-down configuration, the main communications lasers on the instrumentation side will be facing the Earth, enabling high-bit-rate communications without exposing the instrumentation and electronics to ISM wind damage. Note that the first probe launched will not have to enter a face-down configuration, and the later launched probes will advance to join it.
    \item 
    With an edge-on versus face-on differential  thrust $\sim$4 $\times$ 10$^{-3}$ m s$^{-2}$, the swarm assembly time is $\lesssim$2.79 years, with the last probe launched taking the longest time to assemble. As they reach the swarm assembly point, the swarm members will resume an edge-on flight configuration to reduce both risk and drag.    With six intra-swarm communications lasers per probe, a hexagonally symmetric ``snowflake'' configuration will be adopted to optimize intra-swarm communication with an extended swarm.
    \item 
    In the long cruise phase voyage towards Proxima Centauri, the swarm will continue a variety of observations, both regular monitoring of conditions in interstellar space, and more intensive observations of both planned and \textit{ad hoc} targets of opportunity (see Section \ref{sec:science-en-route}). 
    \item 
    During the cruise phase, from time to time individual probes may be turned into a face-down orientation to enable high data rate communications back to Earth. With the higher drag, these probes will rapidly drift out of the swarm, falling behind by 
    losing $\sim$345 m s$^{-1}$ after 1 day of face-down communications, and thus subsequently falling behind by $\sim$30,000 km day$^{-1}$. Such probes may be effectively lost to the swarm, or the entire swarm may increase its drag to slow down and rejoin them. 
    \item
    The majority of the available probes, the main swarm, will  be targeted to pass near the primary target, Proxima b, but sub-swarms can be created to pass near other targets in the Proxima system. We assume here that 10\% of the surviving swarm members will be used to form two such sub-swarms, one each for the planets Proxima c and d; these could communicate autonomously with Earth, or select and share their priority data with the main swarm (see Section \ref{subsec:Swarm-Earth-Comms}). This sub-swarm separation will need to start (for the Proxima c subswarm) no later than 6 months before the encounter. 
    \item 
    Table \ref{table:approach-and-departure} shows that detailed observations of the Proxima system could start with an approach video beginning roughly 1 week before the encounter. This will be done by having a small number (we assume 3)  of probes turning to have their instrument side facing forward. This will cause increased drag; after one week those probes will fall behind the main swarm by $\sim$700,000 km. Traveling instrument-side forward will also erode the equipment and optics on that side, possibly causing irreparable damage to those probes.
    These probes will not be able to communicate directly with Earth, but their main communications lasers will be facing the main swarm and they should be able to relay their observations to the main swarm for transmission back to Earth.  
    \item  
    The main swarm will continue edge-on during their  pass through of the  Proxima system. 
    Section \ref{sec:arrival-science} describes possible observations during that period; note that different probes will be rotated about the velocity vector to point at targets of interest (e.g., at Proxima b), which may interrupt some intra-swarm communications. 
    \item 
    ``Contingency sample'' observations will be collected by selected swarm members, which will then rotate to have their instrument-side facing Earth to send back data during the Proxima transit, in case dust or other issues in the Proxima system disrupts or destroys  the swarm. 
    \item 
    After leaving the Proxima system, the swarm will rotate to a face-down configuration to enable high data rate communication with Earth. As this will take $\sim$17 minutes (Table \ref{table:ISM-drag}) by this point the swarm will be at least 0.4 au outbound from Proxima and its inner planets. This will also enable observations and a departure video as they leave the Proxima system.
    \item 
    The swarm will be inevitably dispersed somewhat by the Proxima system encounter,  and will have to reassemble itself as it leaves the Proxima system. During this period the data rate back to Earth will increase as the swarm is reassembled. 
    \item 
    Communication of data back to Earth is assumed to require at least 1 year in a face-down configuration, eroding the solar-sail side of the probes. After this period the swarm will be available for post-encounter cruise phase science; we regard planning for that period as outside the scope of this document. 
\end{itemize}

\section[The Alpha Centauri System]{The $\alpha$ Centauri System}
\label{Sec:Centauri-System}
%% Note - no greek in section titles - it messes up TOC! That's why there is an alternate. 
%%
The hierarchical triple-star $\alpha$ Centauri system, the closest stellar system to the Sun, includes $\alpha$ Centauri A and B (Cen-A and Cen-B), an inner binary of two solar type stars, and the outlying tertiary member, Proxima Centauri (Proxima),  the closest neighboring star to our Sun.
This system provides  excellent targets for the initial exploration of interstellar planetary systems, as it combines both a binary and a single star system, with both solar type stars and a red-dwarf.

\subsection{AB Centauri}
\label{subsec:AB-Centauri}

Table \ref{table:Centauri_System} provides details on the Centauri stellar system and Table \ref{table:Centauri_Planets} provides details on its single known candidate planet. 

Accurate determination of the age of stars generally requires knowledge of their mass. There are accurate astrometric determinations of the mass of Cen-A and Cen-B, and thus there are a number of age determinations for these stars, with recent determinations ranging from 4.85 to 6.52 billion years using asteroseismology, gyrochronology and chromospheric activity limits, while a recent combined inverse mapping method provided an age range of 7.2 - 7.8 Gyr \citep{Thevenin-et-al-2026-a}. 
It seems reasonable to conclude that the AB Centauri system is probably older than the Sun, possibly by as much as 2 billion years, so any biology or technological civilization associated with it could be older than ours, or even now extinct. If Proxima formed with the AB system, it will of course share the same age; if Proxima was captured subsequently, as is quite possible \citep{Feng-Jones-2018-b}, its age is very poorly constrained.  Determination of the age of Proxima Centauri from observations of the star at encounter is therefore a high priority science goal for any initial Proxima mission.

$\alpha$ Centauri A may have a Neptune-mass planet, currently called Candidate 1, in its HZ, but this is at present unconfirmed \citep{Wagner-et-al-2021-a}; it is still not  known if Cen-B possesses any planets at all. \cite{Zhao-et-a-2018-a} used radial velocity data to set upper limits of 53 and 8.4 M$_{\bigoplus}$ for planets in the HZ of Cen-A and Cen-B; Cen-B probably does not contain a HZ Neptune, but could still contain an Earth or super Earth.

\subsection{Proxima Centauri}
\label{subsec:Proxima-Centauri}

Proxima has two confirmed planets with one, Proxima-b, solidly in its HZ \citep{Anglada-Escude-et-al-2016-a}, and the second, Proxima d, being a small body at the inner edge of that HZ \citep{Mascareno-et-al-2025-a}.
More information on Proxima b and Proxima d can be obtained from the Exoplanet Encyclopaedia \footnote{\url{https://exoplanet.eu/catalog/proxima\_centauri\_b--4042/}} 
\footnote{\url{https://exoplanet.eu/catalog/proxima\_centauri\_d--7409/}}, which is regularly updated.
Proxima also has a candidate planet, Proxima c, a possible super-Earth or sub-Neptune well outside of the Proxima HZ \citep{Benedict-McArthur-2020-a}.

Proxima is  currently $\sim$12,950 au from AB, a separation of 2.18$\degree$ in the terrestrial sky \citep{Kervella-et-al-2016-a}. While this separation would mandate separate expeditions for exploration of Proxima and AB Cen, the small angular separation means that the same power beaming and laser communication systems could be used to support near-simultaneous expeditions to both locations. 
Proxima is $\sim$6166 au (0.0975 ly) closer to the Sun than the AB system, and so the differential travel time at 0.2 c is $\sim\frac{1}{2}$ of a year. A slight adjustment in travel speeds or a delay in departure dates would enable expeditions to arrive at Proxima and AB one year apart, enabling the exploration of the entire system in just under 27 years, including travel times and the return of data from Proxima and AB probes back to Earth in successive solar oppositions, simplifying the data return process. 

\subsection{Other Nearby Targets}
\label{subsec:other-targets}

The next nearest star, Barnard's Star, is a galactic halo star at a distance of 5.96 ly, about 40\% further away than Proxima. Barnard's Star
has four confirmed sub-Earth-mass planets (Barnard's Star b, c, d \& e), all of which are hot bodies inside the star's HZ  \citep{Basant-et-al-2025-a}. 

It would require probes $\sim$30 years to reach Barnard's Star at 0.2 c.  More distant stellar systems would of course have longer travel times; the next nearest star with  confirmed planets, Lalande 21185 \citep{Hurt-et-al-2022-a}, is 8.3044 ly distant and would require over 40 years travel time to reach, and almost 50 years between launch and the return of encounter data to  Earth. These systems, plus 
Wolf 359, a system at 7.8558 ly with 1 candidate planet \citep{Bowens-Rubin-et-al-2023-a}, relatively nearby brown dwarf systems such as the Luhman 16  brown-dwarf binary  at a distance of 6.51 ly \citep{Luhman-2013-a} and any nomadic super-Earth mass plants that might be  discovered near the Sun \citep{Lingam-et-al-2023-b}, will likely complete the \textit{in situ} interstellar exploration target list  for the remainder of this century. GJ 887, the next nearest system with a planet in its HZ \citep{Hartogh-et-al-2026-a}, is at a distance of $\sim$10.718 ly and a travel plus data return time of 64 years. A voyage to this star, or any more distant star, would not return data in this century with the laser system parameters assumed here and thus is out of scope for this paper.

\section{The Interstellar Coracle Probe}

In this paper we assume swarms based on large numbers of Coracle laser-sail probes, as described in
Table \ref{table:swarm_parameters}.
Figures \ref{fig:Flyby-1} and \ref{fig:Flyby-2} show artist impressions of ``Coracle'' laser sail probes approaching the planet Proxima b. 
Although laser light-sail probes will of necessity be very thin disks (the current Coracle design is $\sim$10 mm thick), advances in meta-material flat optics will enable high resolution mission imaging and spectroscopy of the target planet. 
The 10,000-gravity launch condition is so extreme that concentrations of mass (``mascons'') must be absolutely avoided, and thus the thickness of the probe must be as uniform as possible.  

\begin{table*}[ht]
  \centering
\begin{tabular}{|c|c|}
\hline
Swarm or Probe Parameter & Value  [units] \\
\hline
Individual probe diameter [mm]                                             & 4000   \\ 
Probe rim height [mm]                                                & 20  \\ 
Main disk thickness [mm]                                       & 10  \\ 
Mass budget: total sailcraft mass [mg] & 3600 \\
Mass budget: total sail mass [mg] & 2600 \\
Mass budget: total payload mass [mg] & 1000 \\
Mass budget: payload disk + apertures [mg] & 330 \\
Mass budget: betavoltaic battery and ultracapacitor pulsed storage [mg] & 330 \\
Mass budget: rim, intra-probe communications, computation and everything else [mg] & 340 \\
Overall input electrical power per probe, at flyby [mW] & 6 \\
Input electrical power to Swarm-Earth or intra-swarm lasers, at flyby [mW] & 4 \\
Optical output power per probe, at flyby [mW] & 0.4 \\
Swarm-Earth communications wavelength source / as received red-shifted [nm]                     & 432 / 539 \\ 
Maximum Number of probe multi-use optical apertures               & 169 \\ 
Intra-swarm (rim) communications wavelength [nm] & 12,000  \\ 
Number of rim transceivers per probe                           & 6   \\ 
Transverse swarm diameter at flyby [km]                                             & 100,000   \\ 
Number of probes at launch & 1000 \\
Number of surviving probes at flyby, after the Proxima encounter (assumed)                                & 300   \\ 
Average probe spacing within main swarm [km]                                             & 6100   \\ 
\hline
\end{tabular}
\caption{Basic parameters of the proposed Coracle probe swarm. The aerographene metamaterial that forms the main probe body has a variable density, tailored for the particular mechanical requirements. A denser layer will support the drive-beam dielectric reflector while the middle layer will be very sparse. There would be thicker areas around the sensor/communications telescope array on the front face and around the betavoltaic, capacitor, and electronics layers to support them. There are 169 hexagon spaces for optical aperture, not all of which may be occupied depending on the mission profile and mass budget.}
\label{table:swarm_parameters}
\end{table*}

\subsection{Material}
We propose to significantly increase the thickness of the sail by utilizing extremely low-density materials to increase the collision cross section of the sail vis-à-vis the oncoming hydrogen flux.  Candidate materials could be aerographene \citep{shah2022synthesis} (density: 0.16 kg m$^{-3}$) and aerographite \citep{mecklenburg2012aerographite} (density: 0.18 kg m$^{-3}$) \citep{behera2021advanced}. To illustrate the superb performance of this material, a 1-mm thick aerographene structure would have an mass-to-area ratio of order of 10$^{-4}$ kg m$^{-2}$.    Due to this exceptionally low sectional density, the performance of aerographene and aerographite for a laser sail is about $10^{4}$ better than for Mylar.
Both materials are completely opaque with an absorptivity of 1, and have already  been synthesized \citep{mecklenburg2012aerographite} in the laboratory, with no apparent roadblocks preventing mass production. These materials thus share many similarities but for simplicity, we will focus on aerographite in this work.

Current structures based on this material are of order 10s--100s of $\mu$m thick `\citep{meija2017nanomechanics}.  Further reducing the thickness to order 10s--100s of nanometers should, in principle, be feasible, as the wall thickness of the graphene tetrapods  is on the order of 10 nm.  The use of smaller tetrapods as sacrificial material on which the graphite is deposited should be possible \citep{meija2017nanomechanics}.  The tetrapod size and shape were based on the t-ZnO and t-AG sacrificial material used for depositing the graphite on those shapes. There do not seem to be principal obstacles to shrinking the tetrapod size to sub-µm to generate a µm-scale porous structure.  While the synthesis seems feasible, the main question is whether the mechanical properties of such a thin, porous structure can satisfy the requirements for an interstellar mission. 

\subsection{Layout and Features}
\label{subsec:layout-features}

The central disk (see Figure \ref{fig:Flyby-2}) consists of a large-diameter (4 m) phase-coherent array of fresnel-like flat optics  arranged in hex-close-packing (hcp, like a honeycomb).  Because they are part of a common monolithic structure, the position of each of these elements known to within a few nanometers relative to the others, which allows for phase coherence of the full optical array, 
enabling the diffraction limit for the beam to be based on the diameter of the array, enabling  sub-arc-second resolutions at optical wavelengths.

The phased array can be used for both imaging the target and emitting coherent photons for communicating back to Earth.  Because it contains up to 169 separate elements, the array will possess a great deal of redundancy. Therefore the array will tolerate a high rate of attrition due the microscopic holes that will inevitably be formed by impacts with neutral hydrogen or helium nuclei at those times when the probe must cruise face-on relative to the ISM.  Although photonic power would be lost with each failure of an annular element, down to a certain level, information would not be lost, like missing dots in an LED traffic light.

As is shown in Table \ref{table:swarm_parameters}, the total probe mass is 3.6 gm, of which 2.6 gm is allocated to the laser sail (which also protects the instrumentation-side when the probe travels face-down, e.g., during the year after the encounter spent transmitting data to Earth). 
The payload thus has a total mass of 1 gm. Of this, 33\% is allocated to the imaging apertures and the disk payload layer, 
 which has a thickness of several hundred microns using a sandwich of aerographene between meta-material films, and 33\% to 
the nuclear battery.
The remaining mass is allocated to the rim, with 14\% allocated to the forward edge ablation shield, a 20 mm high thickened raised rim perpendicular to the disk, and a similar amount to the electronics, computers and lasers for intra-probe communication, which are  concentrated in the 160-degree long trailing edge of the rim. The remaining  6\% of the mass budget is allocated to the movable trim tabs, also located on the rim of the probe. 

Since the probes will fly edge-on without rotation on the roll axis in order to minimize the radiation dose, the leading edge of the probe will consist of a 20 mm thick layer of aerogel faced with a thin sheet of diamond to serve as an sacrificial barrier to absorb impacts from the ISM and dissipate the energy via evaporation.  No electronics are located any closer than 2 cm deep to the outer surface of this barrier, nor do any instrument ports penetrate the surface.  The barrier extends around 200 degrees of the probe's circumference.  No functionality is contained within this buffer except the single layer of electronics buried 20 mm deep under the face of the leading edge which helps assure overall connectivity.

\subsection{Imaging with Folded Optics}
\label{SubSec:Imaging}

\begin{figure}
\centering
\includegraphics[width=\columnwidth]{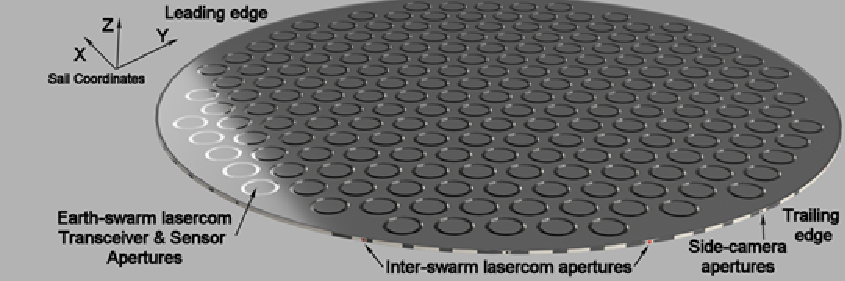}
\caption{Oblique view of the top/forward of a probe (side facing away from the launch laser) depicting array of phase-coherent apertures for both imaging and for sending data back to Earth.}
\label{fig:ProbeFwdHCP-IsoView}
\end{figure}

Although laser light-sail probes will have to be very thin disks (the current Coracle design is 10 mm thick), advances in meta-material optics will enable high resolution mission imaging. 
 Each 200-mm element consists of an annular aperture above an optical well containing a central sensor. 
Figures \ref{fig:ProbeFwdHCP-IsoView} and \ref{fig:ProbeFwdHCP-OpticXSect} show the current design where the probe disk consists of a large-diameter (4 m) phase-coherent array of metamaterial flat optics with 169 smaller 200-mm annular apertures \citep{Skidanov-et-al-2020-a,Park-et-al-2023-a,Eubanks-et-al-2023-b}.

\begin{figure*}
\centering
\includegraphics[width=\textwidth]{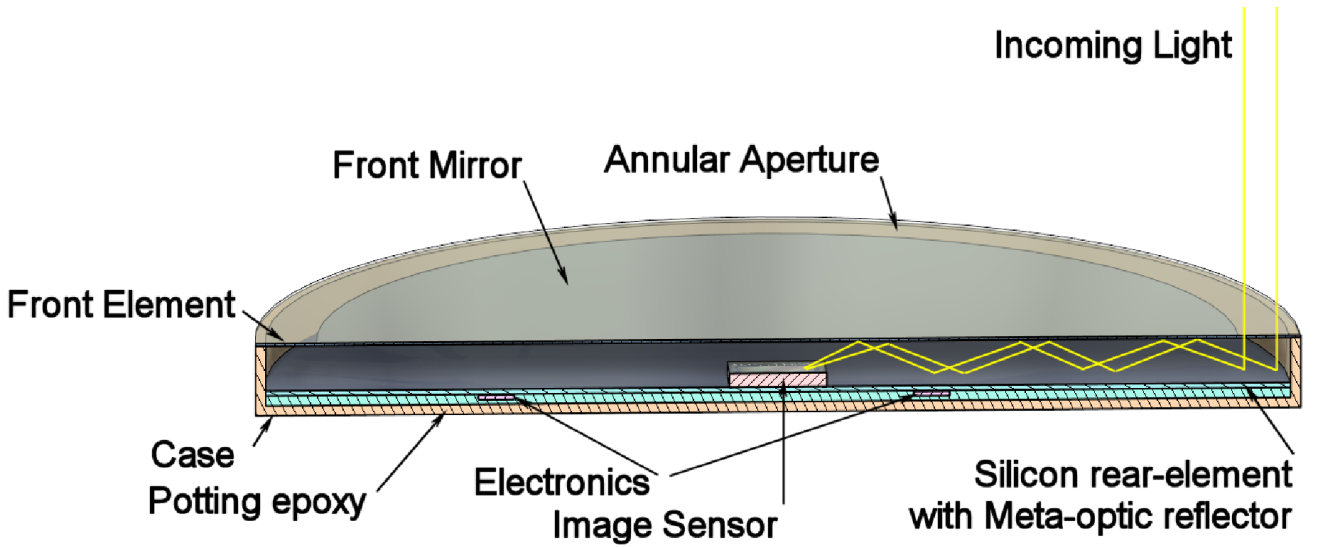}
\caption{The folded optics use an exterior annulus with a width of 25-mm, with internal folded metamaterial reflectors sending light from an annular aperture to an optical well containing a central sensor. Folded optics will be crucial for obtaining high quality images with Coracle mass probes.}
\label{fig:Optics_Cross_section_2}
\end{figure*}

 Figure \ref{fig:Optics_Cross_section_2} shows the 200 mm diameter  folded-optic imaging annulus set into the probe disk \citep{Eubanks-et-al-2023-b}. 
 With each annulus having an area consisting of 5\% of the area of a 200-mm disk, a single disk has 
 a collecting area of 1.57 $\times$ 10$^{-3}$ m$^{2}$, roughly the collecting power of a 2.2 cm disk. If 
 every possible optical aperture is used together the total light collecting area for one Coracle is approximately that of a 0.5 meter telescope.
 While both ISM erosion and image smearing can degrade imaging \citep{Lubin-2016-a}, we anticipate these systems will enable sub-arc-second resolution imaging and spectroscopy of the target planet.

\section{The Proxima Mission: Operational Considerations}
\label{Sec:Operational-Considerations}

In order for an interstellar mission to be successful, various operational issues will have to be resolved; this section presents our list of some of the most important operational considerations for a Proxima Centauri mission.

\subsection{BetaVoltaic Power}
\label{subsec:power}

Betavoltaic nuclear batteries, with energy densities orders of magnitude greater than those achievable with any known chemical method or in-flight electromagnetic or photonic method, 
can carry, in a compact form, sufficient stored energy onboard
to power computation and communication for the entire mission within the rigorous mass constraints of interstellar flight \citep{Dixon-et-al-2016-a,Zhou-et-al-2021-a,Agyekum-2025-a,Gunning-Kennedy-2025-a}.  The selected betavoltaic source material, $^{90}$Sr, is many orders of magnitude cheaper than any other candidate, can be made radiation resistant \citep{Lei-et-al-2020-a}, and could attain commercial-off-the-shelf (COTS) status within a decade with the right incentives. 

The leading edge of each probe rim will be heated by the ISM wind, producing a few mW of thermal power. While this heating may not be useful as a power source, it does depend on the density of the ISM; monitoring it on a minute by minute basis will yield ISM density estimates with a spatial resolution of order a few million km throughout the entire mission, a very useful scientific measurement for studies of the ISM (and the Proxima system stellar wind).

\subsection{Swarm-to-Earth Communication} 
\label{subsec:Swarm-Earth-Comms}

We assume interstellar communications with
2-symbol Pulse Position Modulation (2-PPM), which is widely used in optical communications.  2-PPM uses synchronous time slots, with two adjacent time slots for each bit. A ``0'' value is sent with a pulse in slot 1 and no pulse in slot 2; and \textit{vice versa} to send a ``1'' value.  Symbols are transmitted at 1000 Hz, with pulses nominally 1 ns in duration and the integration slots 10 ns in length.   This technique also has the desirable effect of greatly lowering the unwanted background noise, since the integration time is very short. (Higher order PPM symbols may be used, but for simplicity, we do not consider these here.)  

Given a final velocity of 0.2 c, a transmitted wavelength of 432 nm (blue), is Doppler-shifted to 539 nm (green) when received at Earth.  
The traditional link-budget algorithm used for optical communications is
\begin{equation}
\label{Power}
P_{r} = P_{t} G_{t} G_{r} L_{s} L_{a} \eta_ {pt} \eta_{t} \eta_{r} ,    
\end{equation}
where 
\begin{itemize}
    \item P$_{t}$ is the transmitted power; 
    \item G$_{t}$ is the transmitter gain; 
    \item G$_{r}$ is the receiver gain; 
    \item L$_{s}$ is the path loss; 
    \item L$_{a}$ is the atmospheric loss; 
    \item $\eta_{pt}$ is the pointing efficiency; 
    \item $\eta_{t}$ is the transmitter efficiency and
    \item $\eta_{r}$ is the receiver efficiency. 
\end{itemize}
The transmitter and receiver gain are calculated from the diffraction limits of the telescope apertures for the laser beam wavelength and are expressed relative to an omnidirectional antenna \citep{Marshall1986ReceivedOP, Yuen2022, Wang-et-al-2014-a}. 

For each probe, the electrical power is provided by a betavoltaic battery generating a total of $\sim$9 mW at launch, decaying to 6 mW at the time of flyby, of which a maximum 4 mW is partitioned to the main laser communications system.  This is converted into 0.4 mW of optical power, assuming the current 10 percent electric-to-photonic conversion efficiency.  (We expect this efficiency to improve dramatically over the next few decades.)  Electricity can be stored in a rapid-discharge ultracapacitor.  By concentrating power into 1-ns pulses, the average power of each laser pulse is 400 W, containing 0.4 micro-joules per pulse.

With 270 surviving probes in the main swarm configured  to transmit so the combined pulses arrive together  within a 1 ns window at a desired reception point in the solar system, and a nominal pulse rate of 1200 Hz, each transmitted pulse would yield 3.22 received photons on average, with  an average background level of less than one noise photon per slot, assuming a total reception area of 1 km$^{2}$ \citep{Parkin-2018-a}. Assuming a conservative 1 background photon per slot, and Poisson arrival statistics, this would cause a bit error rate of 0.108. We assume that 20\% of the received symbols are used for error correction, yielding a goodput bit rate of 0.9 kbps and a predicted  Photon Information Efficiency (PIE),
the number of bits conveyed by each received photon, of 0.259. We regard this data rate and PIE as conservative; it is comparable to the PIE of spacecraft laser communications tests to date, but is much below the current experimental record of 14.5 bits per incident photon  \citep{Dacha-et-al-2025-a} or the Gordon-Holevo (GH) capacity bound \citep{Jarzyna-et-al-2024-a} for laser communications.
In conclusion, 0.9 kbps data transfers and  a data return rate of 3.38 Gbytes/year, comparable to the data return of  the \textit{New Horizons} spacecraft after its Pluto flyby, should be possible with the main picospacecraft swarm at Proxima Centauri, and the two sub-swarms should each be able to return $\sim$188 Mbytes/year back to Earth.

\subsection{Intra-Swarm Probe-to-Probe Link Budget}
\label{subsec:Intra-swarm-comms}

The intra-swarm communications SNR  is calculated using the link-budget method; the intra-fleet transmissions are in the long-wave infrared (IR), with a laser wavelength of 12,000 nm. The equations are the same as for the swarm-Earth budget, but only one transmitter and one receiver are used at any one time, and of course the link parameters are different. The intra-swarm transmit/receive apertures are 20 mm diameter flat-optic devices spread around the back, or trailing, half of the outer rim of the probe, which flies edge-on most of the time to minimize radiation dose and erosion by particle flux induced by the ISM at 0.2c. We assume that the optics are electronically steerable, so that high gain may be achieved. Again, the beam is collimated to the degree allowed by the diffraction limit of the aperture.

Given a total of 270 surviving probes with a main swarm  diameter of 100,000 km, and assuming that the probes are evenly spaced, the average distance between probes will be about 6100 km. With a planar swarm in the beta-plane (i.e., orthogonal to the swarm's velocity vector), neither the Sun or  the Centauri system stars will be in the field of view of the intra-swarm transceivers, and their background is assumed to be the general sky noise used by astronomy.   If we allow the maximum high bit rate path length to be triple this, so that each probe can talk to several neighbors, then at this distance the high-rate intra-swarm communication link provides 2.5 photons per time slot, a SNR of 27 dB and a bit rate of 10 kbps \citep{Eubanks-et-al-2023-a}. The ratio of this intra-swarm data rate and the Earth transmission rate, 10 in this case, is the minimum we assumed would be adequate for distribution of data from the probes acquiring or computing it to the complete swarm.

\subsection{Establishing a Coherent Swarm}
\label{subsec:Swarm-Coherence}

A swarm of probes has a coordination problem after it is launched---at first, its members will not on their own know where the other probes are. As described in the Conops (Section \ref{section:ConOps}), we have developed preliminary protocols to develop swarm coherency, defined as a set of probes with intra-swarm communication, positional knowledge, and the ability to position swarm members in a desired configuration. We find that a swarm with a diameter of $\sim$100,000 km can, with assistance from Earth, gain ``self-knowledge'' and   configure itself in deep space \citep{Eubanks-et-al-2023-a,Ding-et-al-2023-a,Dennison-et-al-2023-a}. Establishment of coherence  
 will occur in several distinct phases (see also Section \ref{section:ConOps}):

\begin{enumerate}
\item \textbf{Discovery} Like the synchronous fireflies of the Great Smokey Mountains, swarm members have to find and establish communications with each other.   We assume that the swarm have been placed into a elongated string by the drive process, which is now over. As part of this early stage, the swarm members will change their orientation to be traveling edge-on (See Subsection \ref{subsec:damage-reduction}).
Also, in the first month after launch the swarm will be within a light week (1000 au) of Earth, can be tracked from Earth with mas precision (a 899 km transverse error  at 1000 au) and can also be individually ranged from Earth; this Earth-based positioning information can still be relatively rapidly transferred to the swarm. As this information is received, the swarm will have to configure its drag flaps to bring the swarm members physically close enough to detect each other and begin mutual communications.
The probes have multiple 1-mW pulsed IR transceivers around the rim with 20-mm flat lenses and quantum dot lasers, for sending and receiving to other members. In this period each probe will be placed into slow rotation about its line of flight, in order to search for other probe members. 

\item \textbf{Probes as Beacons} 
During the initial link-up phase, the probes will have to find each other to form a swarm. Initially. accurate locations for the entire swarm can be uplinked from Earth, but the probes are subject to drag, which will perturb their trajectories. As the one-way delay from Earth increases, they will increasingly  have to locate each other for intra-swarm communications using internal resources. For this, we assume that the intra-swarm communications receive optics can be adjusted for a very wide field of view of up to 45 degrees and that the transmit beam is formed into a thin vertical fan, several arc-seconds wide by 45 degrees high. Then, as the probe rotates edge-on during cruise (the optics are along the bottom half of the rim, over a span of 160$\degree$) each probe transmits a pulsed beacon while simultaneously scanning for its neighbors. 
The output power estimate assumes
 that the generated power of 4 mW is stored in a super-capacitor and available for laser pulsing.  Any one probe should be able to detect another probe at a range of $\sim$1 million km.

\item \textbf{Convergence}. Once the probes are in intra-swarm communications range,  a mesh network can be established using the Mobile Ad Hoc Networking (Manet) protocol while the probes are being maneuvered into the desired swarm configuration.
 The goal of this convergence is to establish mesh communications between all nodes in the swarm, and to optimize both communication and scientific data return by establishing a desired swarm configuration. As the exact number of probes surviving will not be known in advance, a snowflake pattern with hexagonal symmetry will be adopted for the swarm so that its configuration can be adopted to the actual size of the swarm, and modified as need be as probes are lost in flight. 
Eventually, a true distance vector ``map'' of the visibility of other nodes can be iteratively established by each node, determining the true relative position vectors between each swarm members, and including auxiliary information such as the probe status and link SNR, and can be transmitted throughout the swarm so that it becomes aware of its overall status. 

\end{enumerate}

\subsection{Damage Reduction by Traveling Edge-On}
\label{subsec:damage-reduction} 

The VoT-Attitude Adjustment method could be initially be under control from Earth, but soon (due to communication latency), it has to become fully autonomous, i.e., under the control of individual probes and eventually that of the swarm as a whole, which in effect would eventually create a ``hive mind.''  With virtually no mass allowance for shielding, traveling edge-on is the only practical means to minimize the extreme radiation damage and erosion induced by traveling through the ISM at 0.2 c.  Moreover, lacking the mass budget for mechanical gimbals or other means to point instruments, controlling attitude and rate changes of the entire craft in pitch, yaw, roll, is the only practical way to aim onboard sensors for image acquisition, and lasers for intra-swarm communications and  interstellar communications with Earth. This would be accomplished using 100 mm $\times$ 100 mm movable trim tabs, the only moving parts on the entire Coracle. 

An initial operational objective (see Section \ref{section:ConOps}) is to  converge  the probes into an operational swarm though Time on Target (ToT) and Velocity on Target (VoT) techniques, a combination of coarse control imposed by the launch laser followed by fine control by interstellar sailing,  dissipating some of the velocity of leading probes by adjusting their attitude and projected area in  the oncoming ISM wind so that the hindmost members catch up with but do not overtake the leading members. We anticipate that most of this convergence will be done by alternating between face-down and edge-on attitudes, which maximize and minimize drag, respectively (see Section \ref{section:ConOps}). Cross-range motions, with velocity adjustments up to the order of 1 km/sec transverse to the direction of travel, can also be obtained either trim-tabs or through tilted orientations. 

%%%%%

%%
\begin{table*}[ht]
    \centering
    \begin{tabular}{|c|c|c|c|c|c|c|}
\hline
     Star    & Spectral type & Mass & Radius & Luminosity & Distance & Planets\\
\hline
      Prox Cen & M5.5-Ve & 0.1221$\pm$0.0022 M$_{\bigodot}$ & 0.1542 $\pm$0.0045 R$_{\bigodot}$ & 0.00157$\pm$0.00002 L$_{\bigodot}$ & 4.2465$\pm$0.0003 ly & Prox b, c, d\\
    Cen-A & G2-V & 1.1055$\pm$0.0039 M$_{\bigodot}$ & 1.2234$\pm$0.0053 R$_{\bigodot}$ & 1.5059$\pm$0.0019 L$_{\bigodot}$  & \multirow{2}{7em}{4.344$\pm$0.002 ly}  &  Candidate 1\\
    Cen-B & K1-V & 0.9373$\pm$0.0033 M$_{\bigodot}$ & 0.8632$\pm$0.0037 R$_{\bigodot}$ & 0.4981$\pm$0.0007 L$_{\bigodot}$  & & none known \\
    \hline
    \end{tabular}
    \caption{The $\alpha$ Centauri System; masses and radii are from \cite{Kervella-et-al-2016-a}.
    The inner AB Centauri  binary consists of two stars similar to the Sun in mass and luminosity
in an elliptical orbit with  a semi-major axis of 23.299 au, an eccentricity of 0.51947 and  a period of 79.762 $\pm$ 0.019 years. From Proxima, 
A and B Cen combined have an apparent visual magnitude of -6.9, considerably brighter than the peak brightness of Venus as seen from Earth. 
Proxima's orbital period is $\sim$550,000 years and it is currently separated from the AB Centauri barycenter by  12,947 $\pm$ 260 au \citep{Kervella-et-al-2016-a}. 
}
    \label{table:Centauri_System}
\end{table*}

\begin{table*}[ht]
    \centering
    \begin{tabular}{|c|c|c|c|c|c|c|c|c|c|}
    \hline
     Planet & Star  & Mass & Radius & Orbital  & Semi-major & m$_{V}$ & Hill Sphere & In HZ? & Confirmed?\\
     &   & M$_{\bigoplus}$ & R$_{\bigoplus}$  & Period (days) & Axis (au) & at SMA &  Radius (km) & & \\
    \hline
    Prox d & Proxima & $\ge$0.26$\pm$0.05   & 0.61  & 5.122$\substack{+0.00019 \\ -0.00022}$  & 0.02885$\substack{+0.002 \\ -0.0036}$ & -26.3 & $\gtrsim$2.6 $\times$ 10$^{4}$  & N? & Y\\
    Prox b & Proxima & $\ge$1.07$\pm$0.06  & 1.01  & 11.1868$\substack{+0.0029 \\ -0.0031}$  & 0.0486$\pm$0.0003  & -22.5 & $\gtrsim$1.5 $\times$ 10$^{5}$  & Y & Y\\
    Prox c & Proxima & 7$\pm$1  & 1.6 & 1928$\pm$20  & 1.489$\pm$0.049 & -15.1 & $\sim$8 $\times$ 10$^{6}$  & N & Disputed\\
  \hline
    \end{tabular}
    \caption{Known and candidate planets of Proxima Centauri. 
     The planetary masses are all lower bounds and depend on the currently unknown orbital inclination; the radii are from interpolation of the rocky models in \cite{Zeng-et-al-2019-a} assuming the lower bound mass. 
     m$_{V}$ is the visual magnitude of Proxima Centauri at the distance of the semi-major axis for each body. For comparison, the Sun at Earth is m$_{V}$ = -26.83 while the Full Moon is m$_{V}$ = -12.6 to -12.7.
    HZ is short for a location in the Habitable Zone.
  }
    \label{table:Centauri_Planets}
\end{table*}

\subsection{Astrometric and Spectroscopic Positioning}
\label{subsec:astrometry-spectroscopy}

Astrometry can be used for interstellar navigation by the swarm. Nearby stars will move in the sky as the swarm moves towards Proxima Centauri, and those apparent motions can be used to position the swarm in interstellar space
\citep{Lauer-et-al-2025-a}. The bright star Sirius, for example, would have a total trip parallax of 26.970$\degree$ going to Proxima Centauri. A single imaging aperture can deliver $\sim$0.4 arc-second astrometric accuracy from a single observation, which near the end of the voyage would deliver a radial position determination with an accuracy of $\sim$1.1 au, while a fully phase coherent 4-m diameter aperture array would improve this positioning by a factor of $\sim$20. 

By observing a larger set of observations of  both local and more distant stars (there are 37 stars with optical magnitudes $\leq$ 10 and trip parallaxes $\ge$ 5$\degree$) it should be possible to position the swarm with optical astrometry, totally autonomously, with an accuracy of order 10$^{7}$ km on each component of position using a few hundred sequential measurements. 

With these positioning accuracies
repeated astrometric positioning over a few weeks would not be able provide short term astrometic velocity determinations to much better than 100 km s$^{-1}$, which is not adequate to monitor short term variations in the edge-on drag. Since spectroscopic radial velocity measurements can be made with an accuracy of $\sim$1 km s$^{-1}$ with relatively small telescopes, astrometric positioning could be combined with spectroscopic radial velocity measurements to measure the swarm's position and velocity on a regular basis. 
The combined data would be adequate to  provide monthly-average measurements of the drag even for probes traveling edge on. This would, together with measurements of rim heating (see Subsection \ref{subsec:power}), provide both unparalleled ISM density estimates and accurate swarm positioning.

\subsection{Interstellar Ranging and Time Transfer}
\label{subsec:interstellar-ranging}

The swarm and Earth should be able to maintain two-way communication throughout the Proxima mission. This will enable both  ranging and time transfer over interstellar distances; this will be important both for the scheduling and timing of mission operations, and for a variety of scientific measurements. 
Given that
the probes will be traveling at a substantial fraction of the speed of light over varying gravitational potentials, including those from gravitational radiation, a  rigorous ranging and time transfer model will have to be fully relativistic, including a proper general relativistic model \citep{Edwards-et-al-2006-a,Messerschmitt-et-al-2023-a}. Here, we will consider various noteworthy effects in a quasi-Newtonian framework. 

With the continuing improvements in atomic clocks, we assume that the probes will use miniaturized atomic clocks with a long term fractional frequency stability (FFS)  $\sim$10$^{-18}$,
comparable to the FFS of modern optical atomic clocks. While this is better than the FFS of atomic clocks now in space,  $\sim$10$^{-16}$, this goal seems reasonable for future single ion optical atomic clocks
\citep{Delahaye-Clement-2018-a}.
If this timing accuracy could be maintained over the full 21 year voyage, the integrated clock error at  arrival at Proxima would correspond to $\sim$0.7 nanosecond, a nominal one-way range error of $\sim$20 cm. Of course, clock comparisons will also be limited by the error in determining the swarm ephemeris, which will require a variety of measurements, including two-way ranging with the solar system, and also very good modeling of the ISM drag on the Coracle probes.

The gravitational red shift due to the Milky Way galaxy is relatively large, $\sim$5.8 $\times$ 10$^{-7}$ using the circular velocity model in 
\cite{Eilers-et-al-2019-a}, but this timing effect will not be directly observable. The differential galactic time dilation between the solar system and Proxima Centauri, however, is observable, and will depend on the distribution of dark matter between them. 

If there happens to be no dark matter between the two systems then their galactic orbital velocities would vary according to Kepler's third law, and the Proxima system, being deeper in the galactic gravitational potential well,  would be redshifted by 7 $\times$ 10$^{-11}$. In the more likely case that dark matter is evenly distributed in our galactic neighborhood, with a distribution given by the local circular velocity gradient of (-1.7 $\pm$ 0.1) km s$^{-1}$ kpc$^{-11}$ \citep{Eilers-et-al-2019-a}, then the differential  Proxima redshift would be 8.6 $\times$ 10$^{-12}$, $\sim$8 times smaller. With two-way ranging and clock synchronization, it should be possible to determine this red shift, directly determining the gradient of dark matter in the galactic neighborhood. 

As is described in Subsection \ref{subsec:grav-wave-detection} and shown in F+igure \ref{fig:grav_wave_mass_integration_24},
gravitational radiation is predicted to cause dimensionless gravitational potential changes (and thus clock shifts) at the level of 10$^{-14}$ over multi-year periods and multi-light year distances. Separating these waves from other effects may require comparing results from multiple expeditions sent in different directions. 

The solar gravitational redshift at the Earth is $\sim$10$^{-8}$, while the Proxima redshift at Proxima b is larger, $\sim$2.5 $\times$ 10$^{-8}$.
While these redshifts are relatively large, they will rapidly decrease with distance away from each star. 
At 0.2 c, these two redshift will cause relatively small integrated clock adjustments of  (to first order) $\sim$312 and 
$\sim$47 $\mu$s, respectively, between Earth and Proxima. AB $\alpha$ Centauri will also cause a redshift of $\sim$2 $\times$ 10$^{-12}$ at Proxima, and a much smaller integrated effect. 

One-way ranging from the solar system to the swarm should be possible (given the laser resources available in the solar system) on a regular basis throughout the entire mission.  The one-way range from the Earth (E) to the swarm (S) is simply
\begin{equation}
\label{eq:1-W-range}
    \mathrm{R}^{S-E}(t_{2}^{S}) =  (t^{S}_{2} - t^{E}_{1})  
\end{equation}
where R$^{S-E}$ is the range measurement using a pulse (or other modulation) originating on Earth timed by a clock in the terrestrial time system  and received by the swarm and timed by a clock in the swarm's time system. (It is assumed that the origination time of that pulse by the sender's clock is also transmitted to the receiver.) 
The measured one-way range is thus a pseudorange, a combination of both the light travel time between two events, the pulse leaving Earth and the pulse arriving at the swarm, and the difference between the terrestrial and swarm time scales at the times of transmission and reception. If either the clock difference or  the absolute range is known or modeled, then it can be removed to determine the other term.  
Once the swarm is in the long cruise phase  the clock difference terms in the pseudorange are likely to be slowly varying in time; one-way ranging could thus be use as a biased range measurement, and may prove very useful for monitoring the swarm ISM drag. In addition, as repeated one-way ranging from Earth effectively provides a means of determining when signals sent at a given time from Earth would arrive at the swarm, it should be very useful for predicting the reception of commands from Earth and timing bistatic measurements with Earth. 

Two-way ranging and time transfer requires a two-way transfer of timing information between the participants. As this is more complicated to arrange than one-way ranging, at interstellar distances it is likely to be less frequent, and there may be a considerable delay (possibly days, or even weeks or months) between reception and retransmission of timing messages at the remote end. Given the very long round trip delays in these voyages, round trip measurements will have to be initiated at the swarm to be operationally useful by the swarm. 

Suppose that the  swarm sends a time signal at a time t$^{S}_{1}$ across interstellar space where it is received by the receiver on Earth, E,  at a local time tag t$^{E}_{2}$ and then returned, after a delay $\delta\tau^{E}_{2}$, tagged at the transmit time t$^{E}_{3}$ and measured in the terrestrial timing system. On return the receiver, S (the original sender) should have the time the outbound pulse was sent and the time the return pulse was received, both in the swarm's time system, and 
an estimate or measurement of $\delta\tau^{E}_{2}$ in the time system of the remote site. The two-way range in the sender's time system,  R$_{S}$, is thus
\begin{equation}
\label{eq:2-W-range}
    \mathrm{R}^{S-E-S}(t_{3}^{S}) =  (t^{S}_{3} - t^{S}_{1}) - \gamma_{S}^{E}\ \delta\tau^{E}_{2}
\end{equation}
where $\gamma_{E}^{S}$ is the scaling between the S and E time systems at the time 
$\delta\tau^{E}_{2}$ is measured. $\gamma$ can be modeled from estimates of the relative velocities and the gravitational potentials at the time of reflection of the range signal at Earth. 

As the round trip range is measured by the swarm clock in Equation \ref{eq:2-W-range}, and as on Earth  the local kinematic and gravitational effects can be very accurately calculated, it should be possible to make very accurate round trip ranging measurements, hopefully approaching the nanosecond accuracy level. Interstellar two-way time transfer requires determination of the true one-way delay from a two-way range estimate, and then subtraction of that from the pseudorange measurement in Figure \ref{eq:1-W-range}. While in the solar system it may be possible to estimate the true one-way delay by simply dividing the two-way delay by two, that will not work in interstellar time transfer. The swarm will have to estimate  its ephemeris from ranging and other data, use that to determine the geometric component of the Earth-swarm delay, and then subtract that from the one-way range from Earth. On Earth, these two-way ranges can be made full duplex by comparing two-way ranges initiated at both Earth and the Swarm in the same time intervals. 

The rocky bodies in the Proxima system can be used as a fiducial points for two-way bistatic ranging. 
Suppose that the swarm is making regular ranging measurements with the solar system using Equation \ref{eq:2-W-range}.
As discussed in Subsection \ref{subsec:bistatic-lidar}, the drive laser can be used to send timing signals from the solar system to probes, both directly, and by reflection off of planetary surfaces in the Proxima system. This differential ranging, together with regular two-way ranging with the swarm, would enable the two-way range to be transferred to  Proxima b (or another in-system body) independently of swarm clock  errors. This transfer would, however, be subject to errors in the Proxima b typography, which could easily amount to 100's of meters, and would only be possible for bodies not covered in clouds. 

Determining the  Earth-Proxima range to within even a few hundred m (i.e., to  a fractional error better than 10$^{-14}$)  would  be a major scientific achievement, enabling both fundamental tests of gravity, observations of decadal period gravitational waves, and considerable improvement in  the ephemerides of the Proxima planets, leading to improvements in the targeting of subsequent missions. 
 
\subsection{Proxima Ephemeris Errors}
\label{subsec:Ephemeris-errors}

While a relativistic picospacecraft will be able to do some course correction \citep{Lubin-2016-a}, relativistic terminal navigation should be avoided if possible, due to the lack of time to make significant course corrections, and the need to point probes in specific directions for scientific data collection during the flyby. With $\sim$300 probes, the total ephemeris error needs to be $\lesssim$100,000 km so that at least one probe passes within a diameter of the target planet, allowing for high resolution imagery and also transmission spectroscopy of the planet's atmosphere.  There will be both radial (or along-track) errors and transverse errors in any interstellar mission; meeting this ephemeris error goal will require improvements in both before a Proxima mission is sent.

At a distance of 4.24 ly (Proxima system parameters here are taken from \citep{Faria-et-al-2022-a}) the voyage at 0.2 c would take 21.22 years in Earth proper time. As the data available at Earth is also retarded by 4.24 years (the  ``R\o{}mer delay''), targeting at launch thus requires predictions 25.46 years, or 831 Proxima b orbital periods, into the future. The current Gaia EDR3 \citep{Gaia-2021-a} error in the Proxima Centauri proper motion is $\sim$52 $\mu$as
per year, which amounts to a 260,000 km error over 25.46 years; the angular position astrometry of the Proxima star system is thus by itself close to the needs of an interstellar mission. 

Radial errors might seem relatively benign, but they cause time of arrival (ToA) errors and of course any target is moving during any ToA errors. With the Proxima b orbital velocity being $\sim$47 km s$^{-1}$, radial errors will have to be $<$ 1 au (which would induce a ToA error of 2500 s) to keep the cross-track orbital error in the Proxima b position $\lesssim$ 100,000 km. This will require a parallax accuracy of $\sim$1 $\mu$as or better, a factor of $\sim$50 better than the Gaia EDR3 parallax error.

The orbits of the Proxima planets are not nearly well determined enough for our mission goals. The desired ephemeris error corresponds to $\sim$0.002 of the Proxima b orbit circumference, requiring a fractional orbital period  error at launch of $\sim$2.6 $\times$ 10$^{-6}$;  the Proxima b semi-major axis error thus needs to be at least two orders of magnitude better determined than the current   error of $\sim$43,000 km. Of course, as the inclination of all of the Proxima planets is currently  basically unknown, the other Kepler orbital elements of at least Proxima b will have to both established, and determined to within order 0.1$\degree$ for the angular elements, and  within a factor of 0.002 or better for the eccentricity.

\begin{table*}[ht]
\centering
\begin{tabular}{|c|c|c|c|}
\hline
\multicolumn{1}{|c}{Error}  & \multicolumn{1}{|c}{$\beta$ plane}   & \multicolumn{1}{|c}{Radial} & \multicolumn{1}{|c|}{ToA $\beta$ plane}\\
\multicolumn{1}{|c}{Source} & \multicolumn{1}{|c}{Error} & 
\multicolumn{1}{|c}{Error} & 
\multicolumn{1}{|c|}{Error} \\
%%\multicolumn{5}{|c|}{ }\\
%%\multicolumn{5}{|c|}{ }\\
\hline 
Proxima RA, $\delta$ Errors &  8700 km    &           &      \\
Proxima Proper Motion Errors & 296,000 km  &           &      \\
Proxima Acceleration by AB Cen   & 22,700  km  & & \\
Proxima Parallax Error    &           & 17.447 au  & 1,040,000 km \\
Proxima~b from Parallax Error  &           & 17.447 au  & 2,040,000 km \\
25-year Radial Velocity Error     &        &  0.025 au  &  2900 km   \\  
Proxima~b Mean Anomaly Error   & 7.6 orbits & & \\
\hline 
\end{tabular}
\caption{Current transverse and radial ephemeris errors, together with the ToA errors induced by radial errors. 
The Proxima parallax and proper motion and Proxima~b's orbit determination all need substantial improvement;
Proxima~b's mean anomaly is currently totally undetermined after $\sim$1.7 years. The Proxima acceleration due to the orbit of Proxima around AB Centauri is easy to calculate, but will need to be added to the Proxima ephemeris.
}
\label{table:ephemeris-errors-1}
\end{table*}

\subsection{Pulsar Timing Navigation over Interstellar Distances}
\label{subsec:pulsar-timing}

X-ray pulsar timing has delivered positioning accuracies better than 10 km in solar system tests \citep{Wang-et-al-2023-a}, and either X-ray or radio pulsar timing navigation is a possibility for interstellar travel. However, while pulsar transverse positions are routinely determined to within a fraction of an au with Very Long Baseline Interferometry (VLBI), pulsar radial positions, based on parallax observations, have uncertainties of many ly. A voyage of multiple light years will cause these radial errors to propagate into pulsar transverse position errors, causing arc-second level pulsar angular position changes and pulsar timing errors of  many hundreds of seconds. 

Pulsar distances must thus be much better determined in order to use pulsar timing for interstellar navigation. Fortunately, the first mission (or even a precursor mission) can use the parallax provided by the Proxima-Earth baseline, and direct ranging measurements of that baseline length, to reduce the distance errors of the pulsars observed by up to 5 orders of magnitude, allowing for future interstellar pulsar navigation. In addition, as radial errors also limit the sensitivity of pulsar timing array (PTA) to gravitational radiation, these pulsar range determinations will profoundly improve our ability to monitor long period gravitational waves with pulsar timing. If it is decided to use pulsar timing for the first Proxima mission, a precursor mission or missions (ideally aimed at or near Proxima  or in  antipodal directions) will be advisable to reduce the pulsar radial errors to an acceptable level. Such precursor missions would also provide information on the ISM along to the path to Proxima, which would be very valuable for further mission planning. 

%%%%%
%%%%%%%%
%%
\begin{figure}[tb]
\centering
\includegraphics[width=\columnwidth]{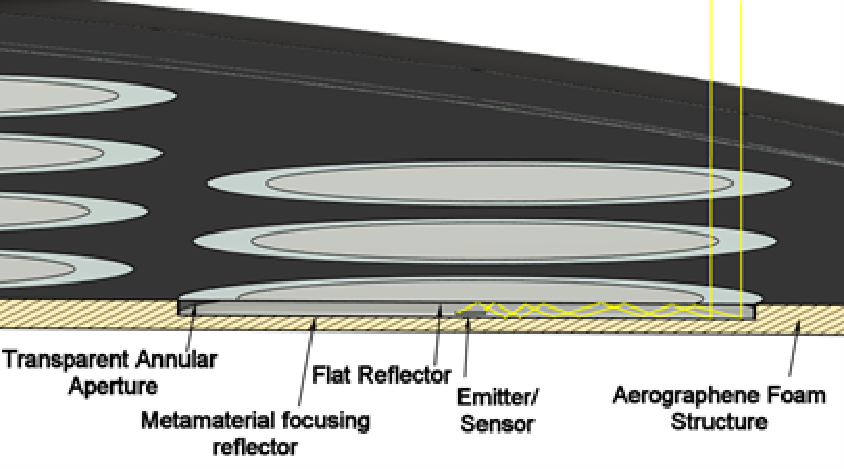}
\caption{Cross-sectional close-up view of one annular aperture in array of phase-coherent elements depicting ray trace from the annulus to the sensor / emitter in the center of the aperture. In this design, light is collected from a total of 5\% of the total aperture area.}
\label{fig:ProbeFwdHCP-OpticXSect}
\end{figure}
%%
%%%%%

\subsection{Gravitational Microlensing Navigation}
\label{subsec:microlensing}

The microlensing rate for a given observer depends on the optical depth of lensing (which depends on the background star field) and the relative lens-observer velocity). The microlensing rate for a relativistic interstellar mission at 0.2 c will thus be $\gtrsim$ 1000 times the microlensing rate at Earth, and the event durations likewise 1000 times shorter, with stellar mass lensing events requiring a month on Earth lasting only about an hour. The terrestrial stellar microlensing event rate, of order 10$^{-7}$ events yr$^{-1}$ star$^{-1}$ \citep{Mroz-et-al-2025-a}, would be accelerated to order 10$^{-4}$ events yr$^{-1}$ star$^{-1}$, suggesting that high-cadence monitoring of $>$ 1000 stars should produce observable microlensing events in a 20 year mission.

It is now possible to predict lensing events for nearby stars \citep{Kervella-et-al-2016-b}.
Probe observations of dozens or hundreds of predicted microlensing events by nearby stars will offer both a means of observing these systems, and a novel means of interstellar navigation. The  probe swarms can also take advantage of their ability to monitor half the sky (see Subsection \ref{subsec:HSA}) to discover microlensing events from  stars and free floating planets  \citep{Sangtarash}. With the microlensing parallax available from the spread of the swarm, at least some of these events could be characterized well enough to determine the mass of the lens \citep{Gould-Yee-2013-a}.  

The speed of the swarm will accelerate the rate, and reduce the duration, of microlensing events by a factor of order 1000. While on Earth stellar-mass lens events can have a duration of up a year and terrestrial-mass lens events typically take a few days or less, the swarm should observe stellar mass events taking less than a day, and Earth mass events taking a few seconds. As the Einstein ring crossing time is $\propto$ Mass$^{-1/2}$ for a given observer, swarm observations also open the potential of observing microlensing by Intermediate Mass Black Holes (IMBH) with masses
$\gtrsim$ 10$^{3}$ M$_{\bigodot}$. These events would take decades or centuries to complete for observers in the solar system, and thus could best be found with high-speed interstellar probes. 

The Sun and the three stars of the Alpha Cen system have each a gravitational minimum focal distance given by $F= c^2 R_*^2/{4GM_*}$ where the Einstein ring of a distant target becomes larger than the stellar photosphere.  For each of the three stars of the Alpha Cen system, one can define a ``focal sphere,'' of radius $F_A$, $F_B$ and $F_C$ respectively. For the Sun and Alpha Cen A and B, $F$ is approximately 548, 742 and 436 au, respectively, while for Proxima $F_C$ is $\sim$107 au. When probes are more distant than the stellar
 gravitational focus, they will also be able to search for planets by microlensing. Similarly, when probes approach the gravitational focal distance of Proxima, Cen A or Cen B. it will cross the focal sphere of one of these stars, where the amplification is maximum.
 
\begin{figure}[hb!]
\centering
\includegraphics[width=\columnwidth]{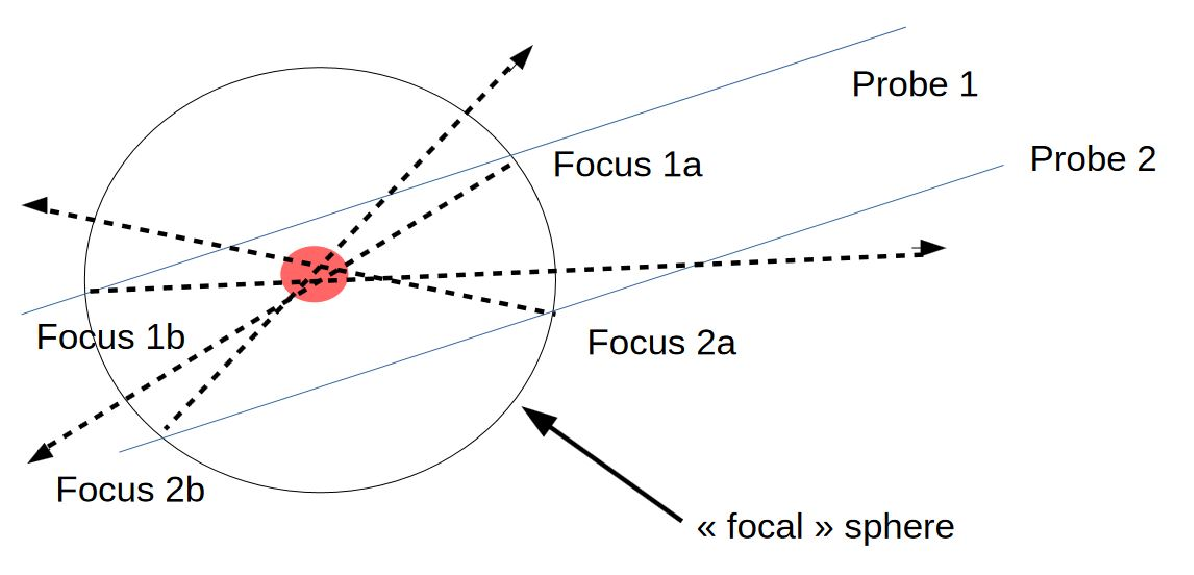}
%%\plotone{Moon_latitude_temp_11.eps}
%%
\caption{Focal points of the paths on two probes. The arrows point toward potential amplified  targets.
\label{fig:Focal-sphere}
}
\end{figure}

\begin{figure}[hb!]
\centering
\includegraphics[width=\columnwidth]{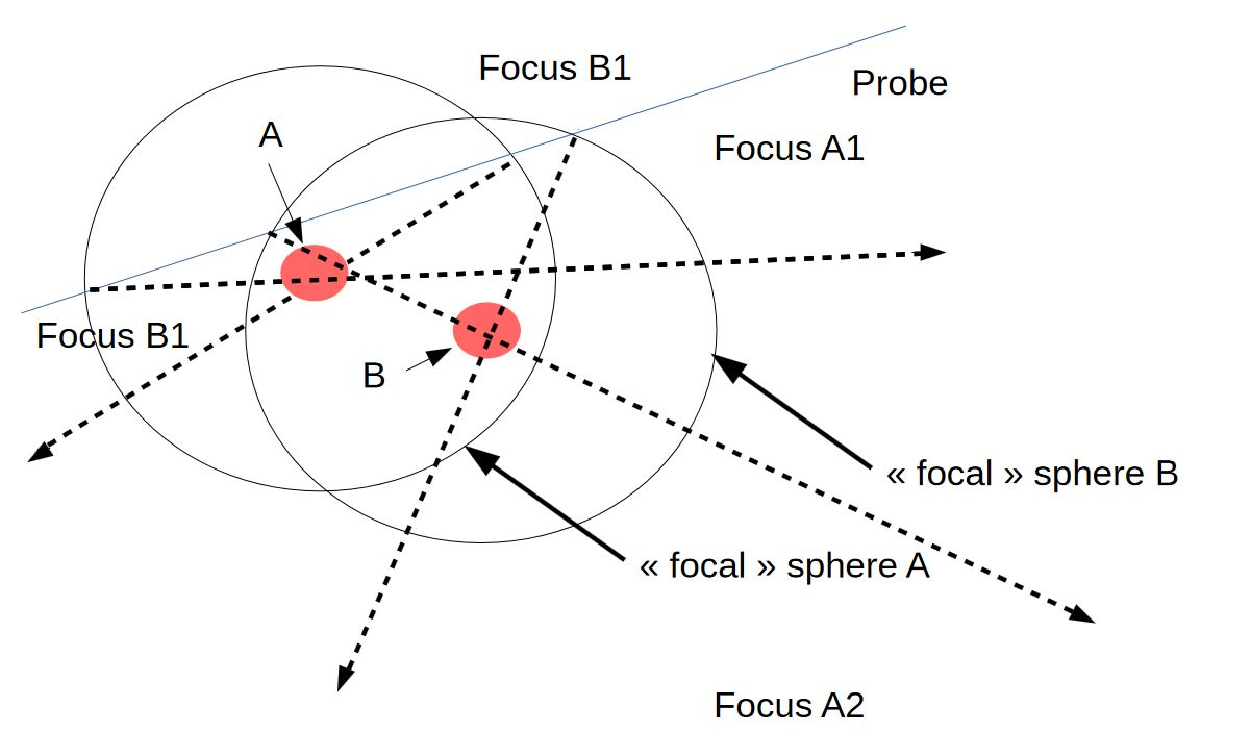}
\caption{Focal points of the paths on one probe crossing two ``focal spheres''. The arrows point toward potential amplified  targets.
\label{fig:two-focal-sphere}
}
\end{figure}

Among the probes of the swarm, a few may have a chance to cross twice one of the three focal spheres. Figure \ref{fig:Focal-sphere} shows the paths of two probes 1 and 2, crossing one of the focal spheres at the focal points 1a, 1b, 2a and 2b (the beginning of the caustic). Each probe $i$ can thus search for planets in the $\vec{*ia}$ and $\vec{*ib}$ directions. Actually, the true picture is more complicated, since the separation between Cen A and Cen B varies between 12 to 36 au, their focal spheres are crossing, as shown on Figure \ref{fig:two-focal-sphere}. 

The focal distances of the Proxima planets are much larger than the Proxima focal distance itself, these being 
0.23, 0.09 and 0.35 ly for Proxima b, c and d, respectively, using the planetary masses and radii in Table \ref{table:Centauri_Planets}. This will enable a ``hybrid microlensing'' zone, where the primary mass (the star) will form a gravitational lens, but the planetary masses will not, although the atmospheres of those planets could form atmospheric lenses, and atmospheric caustics (or mixed gravitational-atmospheric caustics, such  as spikes and a central flash. Atmospheric caustics observed in solar system occultations typically consist of spikes and central flashes, the latter being the atmospheric refraction analog of an Einstein ring passage.
These atmospheric caustics can provide detailed information about even very thin atmospheres
\citep{Pasachoff-et-al-2017-a,Elliot-Veverka-1976-a}, and should be observed if possible during the roughly 1 year period before and after the Proxima encounter when hybrid lensing by the Proxima or AB Cen system is possible. These planetary hybrid lensing events / occultations would of course be very brief for a swarm at 0.2 c, with the entire event lasting for a fraction of a second and any caustics lasting a fraction of a millisecond.


\begin{figure*}
\centering
\includegraphics[width=\textwidth]{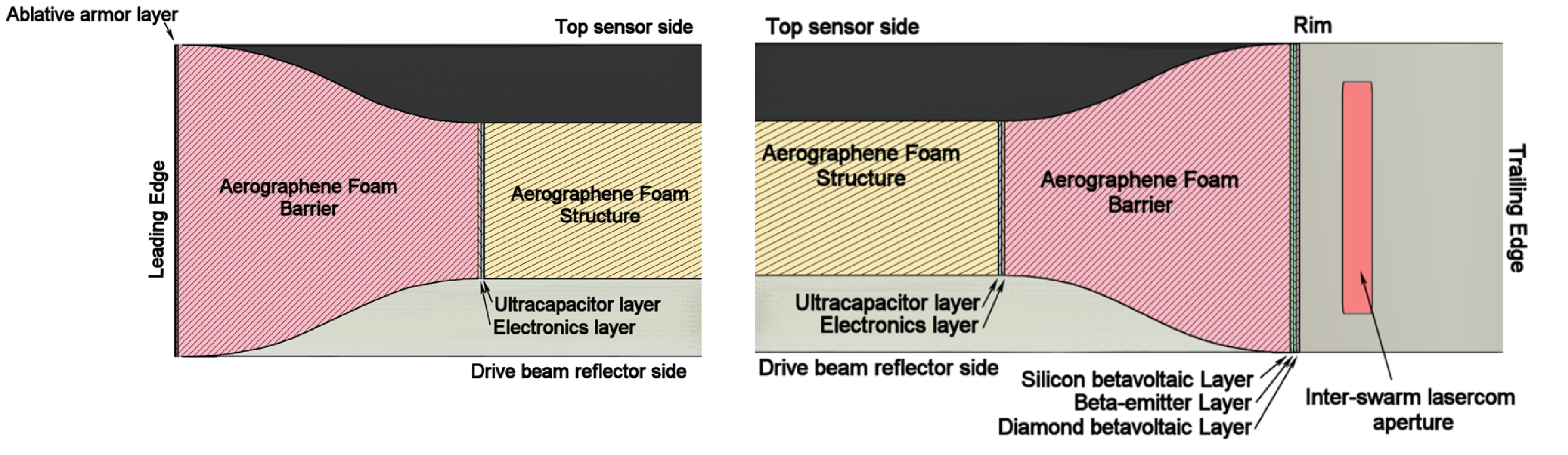}
\caption{Cross-sections of protective leading (to left) edge with sacrificial barrier, and instrumented trailing edge (to right) that contains the betavoltaic battery sandwiches and the probe-to-probe optical transceivers.  Note the electronic layer and ultracapacitor layer span the entire perimeter for mass balance, and to assure connectivity.}
\end{figure*}

\subsection{Interstellar Sailing: Maneuvering in the ISM}
\label{subsec:Interstellar-sailing}

The 4-meter diameter Coracles will normally sail edge-on, exposing a thin armored rim to the erosion of the ISM.
They can, however, act as true sailcraft, with 1000s of au of radial, and tens of au of transverse, motions using ISM drag for course adjustment. Table \ref{table:ISM-drag} shows the  parameters for the ISM drag assumed to be encountered on the trip to Proxima Centauri.   To minimize frontal area, hence dose and erosion, we anticipate flying edge-on most of the way and using the forward edge of the rim for protection against the ISM wind, hence leading and trailing edges are distinct, and configured differently, even though the probe is mostly symmetric.

\begin{table*}[ht]
\begin{center}
\begin{tabular}{ |r|c|c| }
\hline
\multicolumn{1}{|c|}{Parameter} & $\sim$Value & Comments \\
\hline
%%\multicolumn{5}{|c|}{ }\\
%%\multicolumn{5}{|c|}{ }\\
\hline
Collision Rate &   10$^{13}$ p$^{+}$ m$^{-2}$ s$^{-1}$ & or 2 $\times$ 10$^{-14}$ kg m$^{-2}$ s$^{-1}$  \\
ISM Heating    &   2.9 W m$^{-2}$ & Auxiliary Power Source\\
Face-on (FO) drag thrust & 4 $\times$ 10$^{-3}$ m s$^{-2}$ & Launch Flight Mode \\
FO Relative motion in 20 yr & 5000 au & $\sim$0.08 ly  \\
Edge-on (EO) drag thrust & 2 $\times$ 10$^{-5}$ m s$^{-2}$ & Normal Flight Mode \\
EO Relative motion in 20 yr & 32 au &  $\sim$0.0005 ly  \\
Time to Rotate 180$\degree$ & 17 minutes  & with 100 x 100 mm  trim tabs\\
\hline 
\end{tabular}
\end{center}
\caption{Interstellar drag at 0.2 c. Proton collisions are assumed to be absorbed, not reflected or transmitted, with 100\% efficiency. ISM properties are appropriate for the Local Interstellar Cloud (LIC) \citep{Linksy-et-al-2022-a}. With this amount of drag a 1000 probe swarm could coalesce if it was launched over a period  $\lesssim$3 weeks }
\label{table:ISM-drag}
\end{table*}

\subsection{High Dynamic Range Time Delay and Velocity Shift Imaging}
\label{subsec:TDI-VSI}

Imagery during a near-relativistic flyby will be very difficult, as exposures will have to be relatively long, and the targets will both be moving rapidly in the field of view and rapidly changing their kinematic red or blueshifts.  The swarm members will pass by the target planet at a velocity of $\sim$6 $\times$ 10$^{4}$ km s$^{-1}$. At the closest target distance considered here, 1 resolution element ($\sim$20 meters) is traversed in $\sim$0.3 $\mu$s, while comparable imaging of the day-time Earth from Proxima  is conducted with exposures of order 0.01 s. Assuming similar exposure times in a Proxima b, the image smear would be $\sim$30,000 pixels.

Three complementary techniques will be used to improve image clarity,  high-dynamic-range (HDR) imaging, 
Time Delay Integration (TDI) and Velocity Shift Integration (VSI). (HDR and TDI are existing spacecraft technique \citep{Zhai-et-al-2018-a,Guo-et-al-2025-a}, while VSI is apparently a new technique.)
In HDR, multiple images are acquired with very short exposures (e.g., $\lesssim$1 $\mu$s in the Proxima flyby); these underexposed images are registered and stacked with TDI and VSI to obtain the desired total image integration time.
In TDI  severely underexposed short-duration
images are shifted in the image plane as the target moves, while in VSI 
images acquired in different spectral bands at different times are added. Typically both techniques will have to be applied 
 to ensure received photons
for a given frequency band and a given spot on the target
are all integrated on the same image pixel. It will not be possible to predict this registration in advance for all targets of interest, and objects with different speeds or projected locations in a field of view may have to be shifted differently.

In 0.01 s the spacecraft would move $\sim$600 km, which, at a distance of 10,000 km, which would cause noticeable distortions of the images being stacked; these are predictable and can be removed. Iterative HDR can remove rotations of the spacecraft during the image, correct for ephemeris errors during imaging, and also correct smearing due to objects with different relative velocities in the image plane. In a 10 second flyby with 10$^{6}$ mega-pixel images per second per aperture a single probe with multiple aperture arrays might obtain billions of images, mostly greatly underexposed. This will form the raw material for searches for small bodies and unanticipated features in the Proxima system. It will never be possible to send all of this raw material back to Earth; extracting as much useful information as possible from it after the encounter will be a major computational task for the probes in the swarm.

\section{Autonomous Data Selection with data broker agents}
\label{Sec:data-selection}

The close flyby of Proxima b will last less than a minute, with the highest image resolution will come from the small subset of probes that happen to pass closest to targets of interest.
In addition, with 
HDR recordings acquiring up to one million images per second per aperture (see Subsection \ref{subsec:TDI-VSI}) there will be terabytes of data acquired by the swarm during the flyby. The swarm will thus inevitably have orders of magnitude more data than can be sent back to Earth even with years of data download by thousands of probes working together \citep{Eubanks-et-al-2023-a}.  Scheduling of observations, and their ranking and down-selection for transmission back to Earth, will thus be critical part of mission design, especially as these operations that will have to be done without any immediate feed-back from mission control in the solar system. 

\subsection{Dynamic Targeting with Swarm Spacecraft}
\label{subsec:dynamic-targeting}

Dynamic Targeting (DT) is a mission operations concept to use early observational information (``lookahead data'') of a target system to better plan  subsequent observations 
\citep{Kangaslahti-et-al-2026,Candela-et-al-2023-a}. In DT lookahead data acquired before the start of the main observing campaign, or before individual observations, 
can be used to both schedule more images of interesting or dynamically unusual events, and to skip observations that are not likely to fulfill scientific goals. In the Proxima mission this lookahead data will be taken from data acquired during the approach period, the roughly 8-day period just before the encounter (see Subsection  
\ref{subsec:approach-imaging}).

DT will be integrated into agentic system (see Subsection \ref{subsec:sorting-hat-brokers}), both within the domain of specific agents and also at a higher level between agents. Broker agents searching for undiscovered asteroids and moons, for example, could use lookahead observations to find and then schedule imaging of candidate objects. Agents searching for planetary craters could schedule observations of candidate craters, but if the planet proves to be cloud-covered the higher command layer could, e.g., remove observing time from crater imaging and allocate it to planetary meteorology. 
Similar imaging reconfiguration adjustments could also be applied to spectral bands. If, for example, Proxima b is similar to Venus or Titan with a heavy cloud cover in most wavelengths but with spectral windows allowing imaging of the planet's surface in certain wavelengths, then more observing time or bandwidth could be allocated to those wavelengths.

\subsection{Distributed Data Selection with Sorting Hat Brokers}
\label{subsec:sorting-hat-brokers}

With the entire swarm coordinating the return of data to Earth, it should be possible to support a data rate of $\sim$0.9 kbps, or $\sim$3.4 gigabytes per  year (see Subsection \ref{subsec:Swarm-Earth-Comms}). While comparable to the New Horizons data return from Pluto, this implies that it will not be possible to return more than a small fraction of the many terabytes of data likely to be acquired by the swarm flyby of Proxima 4 ly from Earth. The data selection process will thus demand new means of filtering and selecting data for  return  to Earth. Data-broker-agents (or agents)  —  automated software systems designed to support different scientific goals by sifting, characterizing and prioritizing swarm observations - will be critical tools for managing the floods of data from light-sail swarms.

As  inter-probe communication limits the amount of data that can be
shared between probes, and as every probe must share all of the data actually sent back to Earth, data selection must be distributed in nature, and it is not sufficient to simply have each probe select its best image (or other data) to return to Earth. This could create a ``Paris Selfie Problem,'' where an analogous mission sends 1000 tourists
to Paris, gives each a camera with instructions to select their best picture, and
gets in return 1000 selfies of the Eiffel Tower. 
As any interstellar mission will have a wide range of science goals the  data selection system will have to  instantiate the science traceability matrix; an agent based system can allow for autonomous selection of data to best fulfill the multiple science objectives of the mission.

The Roman Space Telescope / LSST has a similar problem in dealing with alerts
for their flood of data, and has solved it by establishing a system of Alert-Brokers
\citep{Narayan-et-al-2018-a}. Part of the power of this system is that different agents can be developed and tuned for different problems, and by different groups, allowing for independent searches for different goals. Based on the alert brokers idea, different agents should be developed for filtering and selection for different purposes, including for biosignatures and technosignatures \citep{Eubanks-et-al-2025-e}.

In the context of a flyby, ``sorting hat'' data-selection-agents should be set up to examine the data stream for different features, based on techniques appropriate for that item. Broker agents searching for the undiscovered asteroids and moons, for example, could be simply searching for point sources moving in a coherent fashion between different images, while other agents could search for specific geological features, such as craters, or even specific technosignatures, with machine learning using extensive libraries of terrestrial images.

A variety of data selection-agents, each focused on a different scientific question
or topic, will be essential for data management with swarm spacecraft  4
ly from Earth. Automatic selection-agents for a given topic can examine, characterize and rank / score data elements for follow-up consideration, but if the swarm is going to cooperate on the transmission of data back to Earth there still has be a means to select that data on a swarm-wide basis. Data agents can also perform the TDI and VSI corrected images for further analysis, and can select and prepare data for other agents. (For example, a agent could find and correct images of the target planet for motion distortion; these corrected images could then be passed to agents searching for planetary features such as craters.) Each agent will be tasked with finding candidate data sets to fulfill its goals, together with a ranking those data sets and also a calculation of the probability of each data set fulfilling its goal. For example, if Proxima b turned out to be a Venus-like planet covered in clouds, an agent tasked with finding images of craters might come up with a ranked list of candidate crater images, but give each a very low probability of actually being an image of a crater. 

Fulfillment of the mission scientific goals will require instantiation of the science traceability matrix (STM) in another set of agents, using weights and procedures largely or entirely set before launch. To continue the example, planetary geology might be given a certain weight in the STM, and craters might have a certain fraction of that, say 1\% of the total weight, but that does not mean that 1\% of the total data return would necessarily be devoted to crater images. If the crater image candidate probabilities are very low, then maybe a minimum number of such images will be returned. Certain STM goals, such as the search for technosignatures, might get high priority immediate transmission if their estimated probability is high, but have to wait their turn if it is not. 
Construction of an agent system to manage the STM goals of the mission (together with other goals, such as returning an adequate amount of engineering data) will clearly require considerable work before the first missions are sent.

\subsection{OESF loops in Swarm data Selection}
\label{subsec:OESF-loops}

During and after flyby data collection the swarm faces a difficult data handling problem. There will be a mass of redundant data collected (nearly identical images of the target planet, for example), and having each probe flood all of its data to every other probe is not an efficient use of intra-swarm bandwidth. This can be solved through the use of 
Observe – Evaluate – Select – Flood
(OESF) Loops, and the division of the swarm into nested sets of nearest-neighbor groups. 

A agent tasked with looking for craters, for example, could in the evaluation stage, examine all of the images acquired by its probe and give them a score as crater candidates. In the selection stage, a small set of nearest neighbors to this probe could then compare their candidate rankings, and derive a candidate winner for the entire neighborhood. That would be repeated for a larger sets of neighbors of neighbors, until the entire swarm has agreed upon a winning candidate, without requiring a wholesale transmission of candidate images. At this point the winning candidate can be flooded to the entire swarm for transmission back to Earth. In addition, each probe will then know the transmission winner, and can avoid selecting a similar image in the next OESF loop. These OESF loops could thus terminate quickly (if no good candidates for a given goal are found) or continue through the time allotted for data return (if there are many such candidates).

While in a planetary flyby the data acquisition will mostly be over quickly, data analysis, combination and selection could continue for most if not all of the one year allocated for data return. This process will inevitably be iterative, as both findings and knowledge of the data already returned will inform later data selections. An agent on one probe tasked with finding craters on Proxima b (as in the previous example) might conclude it has found a good candidate (the evaluation and selection step), and promote the images of that candidate for return to Earth (a flooding step) while also flooding its position on the planet throughout the swarm to enable agents on other probes to search for other images of that crater. Agents tasked with finding orbital bodies (moons, asteroids and comets) could determine initial orbits and share the orbital elements with other probes to enable them to find other images of the same object and use that data to improve its orbit further, which might enable finding even more images of that candidate.

%%%
\begin{figure*} 
\centering
\includegraphics[width=0.8\textwidth]{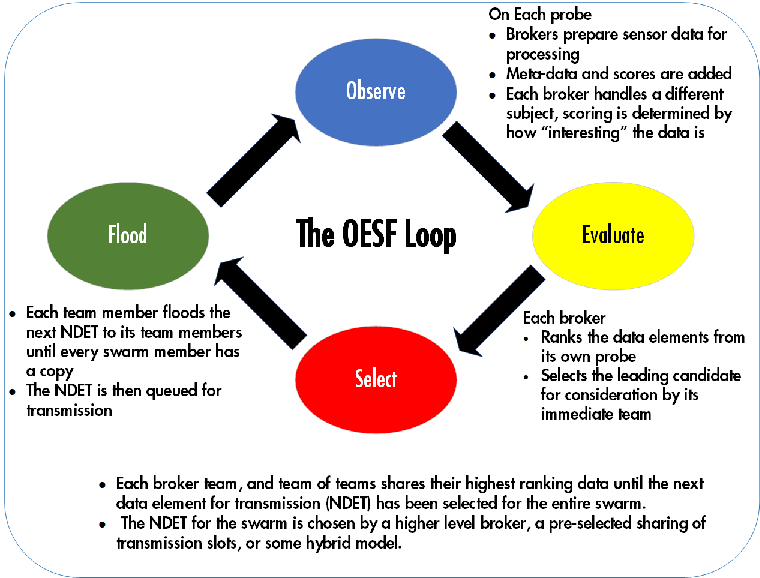}
%%\plotone{Moon_latitude_temp_11.eps}
%%
\caption{The basics of the OESF loop. To minimize bandwidth usage, this would kept to small sets of neighbors, neighbors of neighbors, etc.
\label{fig:OESF-loop}
}
\end{figure*}
%%%%%

\section{En Route Science}
\label{sec:science-en-route}
 
The top side of the swarm spacecraft (the side not exposed to the drive laser) will be covered by instruments and laser and imaging apertures (Figure \ref{fig:ProbeFwdHCP-OpticXSect}.) Lenses or other instruments facing forward will be exposed to the erosion from the ISM and interstellar dust, and would be unlikely to survive intact more than a few months. The swarm spacecraft will thus have to flip after the drive laser thrust period is finished, to protect their  instruments from ISM drag during the voyage.

 With a large swarm, some probes could be sacrificed to turn forwards and observe the Proxima system, especially as the encounter approaches. If probe instrumentation would last 3 months subjected to ISM erosion, continual forward observation during the entire mission would require 84 probes, or a substantial fraction of the total swarm. Forward observations are thus likely to be devoted to observations of the Proxima system, and it  is clear that substantially more observational resources can be devoted en route to the trailing hemisphere.

\subsection{Monitoring the Trailing Hemisphere}
\label{subsec:HSA}

 While some en route observations will undoubtedly be preplanned and focused on targets of interest, it would also be scientifically useful to scan as much of the sky as possible for changes and transient phenomena. The orientation of the swarm will enable observation of objects and events in the swarm's trailing hemisphere during the 21-year voyage to Proxima Centauri. 

In order for the swarm to monitor the entire trailing hemisphere, 
we propose a Half-Sky Array (HSA), built by using four or more existing apertures as wide-angle imaging systems, each looking in a  different direction, with each probe. With (assuming some losses and overlap) $\sim$1000 independent fields of view available, scanning  the entire trailing hemisphere would require each camera to have a field of view of $\sim$2.6$\degree$ which (with megapixel imaging) would provide gigapixel imaging of the trailing sky. This could be used to search for  transient phenomenon during the voyage, and also search for asteroids and comets both as the swarm was leaving the solar system and as it entered the Proxima system. Higher resolution cameras could be autonomously targeted on interesting objects or events found by the wide-angle array.

\subsection{Stellar Occultations en route}

During its two years voyage through the Solar System's Oort Cloud, 
the probe will observe stellar occultations by small bodies.  Beyond the Oort Cloud, the probe will similarly observe stellar occultations by interstellar asteroids or rogue planets, if any, and then there will be a period when it is exposed to the shadows of bodies in the Oort 
cloud of the $\alpha$ Centauri system. 

Stellar occultations observed by a probe swarm will enable occultations that are not visible from the inner solar system. An unexplored area, the occultation of binary stars by asteroids, is in preparation (Souami, private communication). The probe will observe those occultations which are missed from Earth. They will constrain the geometry of unknown orbits, especially for spectroscopic binaries  unresolved by imaging. Similar space occultations have already been observed with the CHEOPS satellite \citep{Morgado-et-al-2022-a}. 

For most of the long journey to Proxima Centauri, the occulting bodies would likely be new discoveries; this thus provides a means of searching for nomadic planets and also for objects in the Oort cloud of both the Solar and Alpha Centauri systems.
With the high probe velocities, an occultation by an Earth-sized body would last no more than 0.2 seconds, and only the brighter stars would support monitoring at the millisecond integration times needed to detect the largest known Oort cloud object,  C/2014 UN$_{271}$, with an estimated diameter of roughly 100 km \citep{Hui-et-al-2022-a}. As with the HSA in general, occultation searches would result in a huge amount of data acquired by the probes, only a small fraction of which would need to be relayed to Earth. 

\subsection{Detection of Nearby Bodies from Trajectory Deviations}
\label{subsec:trajectory-perturbations}

We can investigate how the presence of a single body $X$, with gravitational mass $\mu_X = GM_X$, may through its gravitational influence, affect  the trajectories of $N$ probes as they fly past the object. Recall that the swarm is traveling with close-to-relativistic speed $V_S = 0.2c$, and so we can confidently assume that the speed of the body $X$ is small such that $\lVert \vec{v}_X
 \rVert << V_S$.\\

We find that the mysterious object $X$ will have two enduring and thus measurable manifestations on the paths of these $N$ probes, in addition to one temporary effect which may not be so discernible. The two enduring effects are listed first and second in Table \ref{MANIFEST}, and the ephemeral effect is the third item.

\begin{table*}
\caption{How a single body X might alter the trajectories of N probes formed in a swarm}
\begin{tabular}{ccc}
\hline
\textbf{\begin{tabular}[c]{@{}c@{}}Manifestation on \\  the probes by X \end{tabular}} & \textbf{\begin{tabular}[c]{@{}c@{}}Perm/ \\ Temp \end{tabular}} & \textbf{Description} \\ \hline
\textbf{1} & \ P & \begin{tabular}[c]{@{}c@{}}Each probe, $i = 1,..., N$ will be deflected through\\ a different angle $\alpha_i$ by the encounter, as is manifested\\  by the change in direction it is traveling relative to a fixed\\ inertial reference frame.\end{tabular} \\ \hline
\textbf{2} & \ P & \begin{tabular}[c]{@{}c@{}}Each probe’s change in velocity direction will be\\  such that the common plane shared by the velocities before\\ and after the encounter, $\vec{V}_{i,1}$, $\vec{V}_{i,2}$ will  also contain the object X itself.\end{tabular}\\ \hline
\textbf{3} & \ T & \begin{tabular}[c]{@{}c@{}}As the probes approach the object X, their velocity vectors\\ will increase in magnitude up to periapse, after which\\  a deceleration will return these magnitudes back to their original\\ values, albeit deflected in direction as explained in 1 \& 2 above.\end{tabular} \\ \hline
\end{tabular}
\label{MANIFEST}
\end{table*}

\begin{figure}
\hspace{-0.6cm}
\includegraphics[width=\columnwidth]{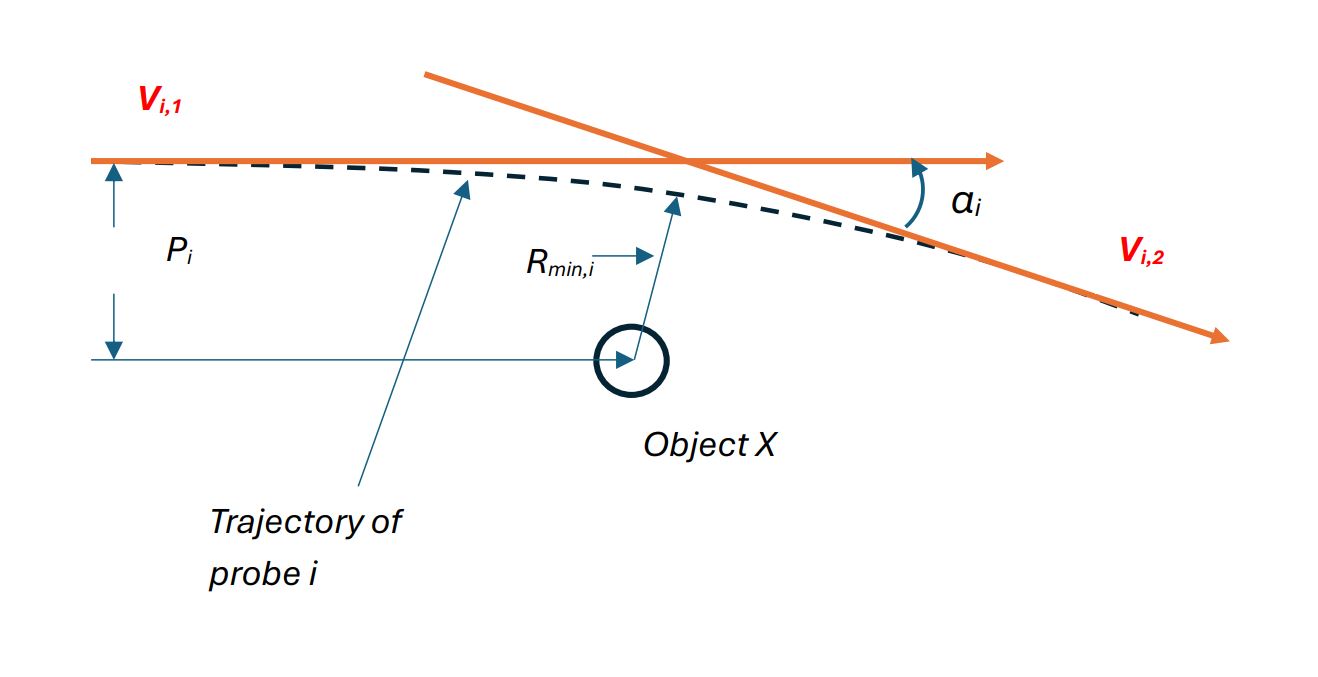}
\caption{Definition of angle $\alpha_i$ for a probe encountering a single body $X$}
\label{fig:Alpha}
\end{figure}

Having delineated these manifestations 1 to 3, we first provide the equations which describe item 1 in Table \ref{MANIFEST}, and is a permanent alteration to a single probe's path as follows.
If the velocity of a probe $i$ is denoted $V_S$ (and is identical for all swarm members), its closest approach to $X$ is $R_{min,i}$, then since the impact parameter $P_i$ will differ negligibly from $R_{min,i}$, we can write:
\begin{equation}
\label{eq1}
    \alpha_i = \frac{2\mu_X}{P_i {V_S}^2}
\end{equation}
where the angle $\alpha_i$ is assumed small. The situation is illustrated in Figure \ref{fig:Alpha}.

Furthermore, once the encounter has elapsed and the probe is no longer inside $X$'s gravitational sphere of influence, $\alpha_i$ is a measurable quantity for each probe given by:
\begin{equation}
\cos{\alpha_i}=\frac{\vec{V}_{i,1} \cdot \vec{V}_{i,2}}{\norm{\vec{V}_{i,1}}\norm{\vec{V}_{i,2}}} = \frac{\vec{V}_{i,1} \cdot \vec{V}_{i,2}}{V_S^2}
\end{equation}

We now address the precise manner by which the object X may temporarily affect the speed of each probe member $i=1,...,N$ as outlined in item 3 of Table \ref{MANIFEST}.
To this end, let $V_{max,i}$ denote the peak speed attained by the probe as a consequence of the gravitational force exerted on it by X. This maximum will necessarily occur at the periapse of the probe with X, i.e. when its radial distance w.r.t. X is a minimum, $R_{min,i}$ in Figure \ref{fig:Alpha}. Thus we have from conservation of energy:
\begin{equation}
\label{energy}
    V^2_{max,i} - V^2_S=  \frac{2\mu_X}{R_{min,i}} 
\end{equation}
If we denote the maximum change in speed affected by the encounter as $\Delta V_i = (V_{max,i}-V_S)$, and also if we note that the left hand side of equation \ref{energy} is the difference of two squares, we can rearrange that equation as follows:
\begin{equation}
\label{DeltaV}
   \Delta V_i = \frac{2\mu_X}{R_{min,i}(V_{max,i} + V_S)} 
\end{equation}
Finally we can simplify equation \ref{DeltaV}, since first $R_{min,i}$ differs negligibly from $P_i$, and second $(V_{max,i} + V_S) \approx{2V_S}$, thus we find:
\begin{equation}
\label{DeltaV2}
    \mu_X = \Delta V_i P_i V_S
\end{equation}
We have already precluded knowledge of the location of X, and also knowledge of $P_i$, thus we are as yet unable to calculate $\mu_X$ by equations \ref{eq1} or \ref{DeltaV2}.

As already stated, we have no direct knowledge of the position of $X$, thus $P_i$ in equation \ref{eq1} is unknown, however we do have item 2 in Table \ref{MANIFEST} which, as shall be shown from the following rationale, can be exploited to derive an indication of the position of $X$, and thus $P_i$ and then in turn $\mu_X$ from equation \ref{eq1} or, if we were able to detect the transient change in velocity, equation \ref{DeltaV2} as required.

We remark as explained in item 2 of Table \ref{MANIFEST} above that there is considerable further information to be gleaned from the plane which is made by $\vec{V}_{i,1}$  \& $\vec{V}_{i,2}$ and this information can be used to ascertain $P_i$.

To this end we now denote the angle made by the unit vector in the direction  $\vec{V}_{i,1}  \times \vec{V}_{i,2}$ (where ‘$\times$’ here denotes cross product), with the vertical unit vector of the swarm plane, $\vec{e}_y$, as $\beta_i$. As stated above the plane made by the vectors $\vec{V}_{i,1}$  \& $\vec{V}_{i,2}$ also contains the object in question, $X$.

\begin{figure}
\centering
\vspace{0.8cm}
\includegraphics[width=\columnwidth]{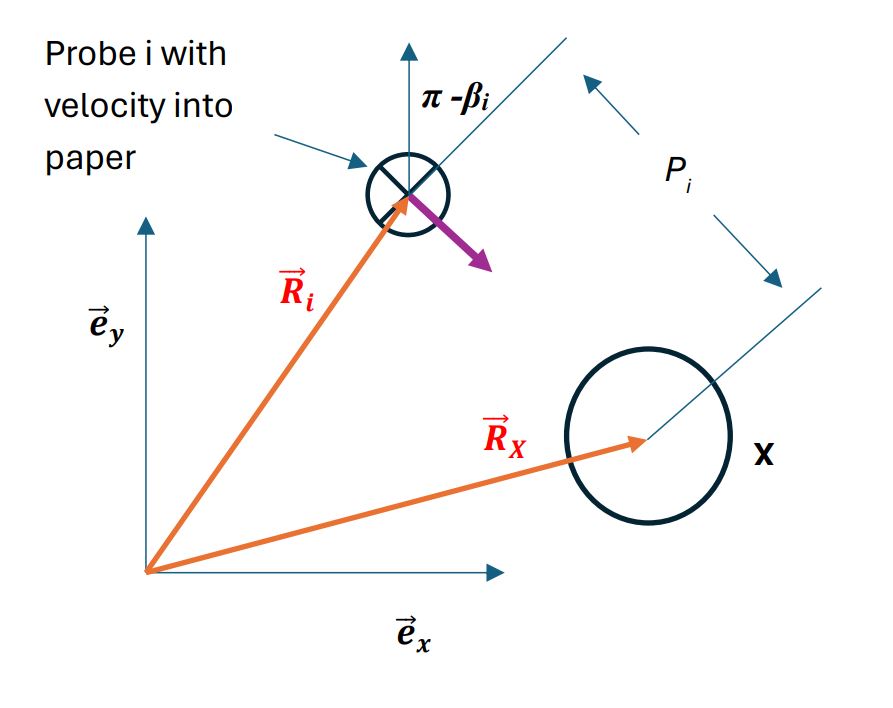}
\caption{Definition of angle $\beta_i$ for a probe encountering a single body $X$}
\label{fig:Beta}
\end{figure}

Thus we have:
\begin{equation}
\sin{\beta_i}\ =\ \frac{(\vec{V}_{i,1}\ \times\ \vec{V}_{i,2})\ \cdot\ \vec{e}_y}{\norm{\vec{V}_{i,1}}\norm{\vec{V}_{i,2}}}\ =\ \frac{(\vec{V}_{i,1}\ \times\ \vec{V}_{i,2})\ \cdot\ \vec{e}_y}{V_S^2}
\end{equation}
Observe that the intersection of the swarm plane (which is defined by $\vec{e}_x$ and $\vec{e}_y$) together with all the planes made by $\vec{V}_{i,1}$\  \&\ $\vec{V}_{i,2}$,\  $i=1,\ \ldots\ N$ (note that all these planes are perpendicular to the swarm plane), marks the placement, $\vec{R}_X$, of the intersection of the body X with the swarm plane. The situation for a single probe $i$ is illustrated in Figure \ref{fig:Beta}.

In this Figure we provide the swarm plane vector basis $\vec{e}_x$ and $\vec{e}_y$, and the position $\vec{R}_i$ in this plane of the $i^{th}$ swarm member (before encounter), together with the previously defined and unknown vector $\vec{R}_X$ denoting the position of $X$. We see that $\vec{R}_X$  lies somewhere along the $\vec{R}_X$ - $\vec{R}_i$ vector in Figure \ref{fig:Beta}, and this arrow is defined by known vector $\vec{R}_i$, unknown scalar $P_i$, and known quantity $\beta_i$ which can be measured after $X$ has exerted its influence on swarm member $i$. 

We have from Figure \ref{fig:Beta} the following:
\begin{equation}
\label{eqRX}
\vec{R}_X =(R_{i,x}+P_i\sin{\beta_i})\vec{e}_x + (R_{i,y}+P_i\cos{\beta_i})\vec{e}_y
\end{equation}
where $R_{i,x}$ and $R_{i,y}$ are the (known) components of vector $\vec{R}_i$.

Reducing equation \ref{eqRX} into its two components and rearranging to get $P_i$ we have:
\begin{equation}
\label{eqPix}
  P_i=\frac{(R_{X,x}-R_{i,x})}{\sin{\beta_i}}
\end{equation}
and
\begin{equation}
\label{eqPiy}
    P_i=\frac{(R_{X,y}-R_{i,y})}{\cos{\beta_i}}
\end{equation}
Equating \ref{eqPix} and \ref{eqPiy} and manipulating we can define the function $f_i$ such that:
\begin{equation}
f_i(R_{X,x},R_{X,y})\ =\ (R_{X,x}-R_{i,x})\ -\ \tan{\beta_i(R_{X,y}-R_{i,y})}
\end{equation}
and that further for the actual values of $R_{X,x}$ and $R_{X,y}$, then 
\begin{equation}
    f_i(R_{X,x},R_{X,y})=0
\end{equation}
It makes sense then, following this logic, that the unknown vector $\vec{R}_X$ must have components $R_{X,x}$ and $R_{X,y}$ which minimize the function, $g$ where:
\begin{equation}
\label{eqg} g(R_{X,x},R_{X,y})=\sum_{i=1}^{N}f^2_i(R_{X,x},R_{X,y})
\end{equation}
Having found this minimum, then we can compute all the $P_i$'s from the $2N$ equations given by \ref{eqPix} or \ref{eqPiy}. Note however there is more information to be extracted, to wit in equation \ref{eq1} there is a proportionality between the known values $\alpha_i$ and the reciprocal of $P_i$, $\eta_i = \frac{1}{P_i}$ as follows:
\begin{equation}
\alpha_i \propto \eta_i
\end{equation}
where the constant of proportionality, $k$, is given by
\begin{equation}
\label{k}
    k = \frac{2\mu_X}{V^2_S}
\end{equation}
Thus if we perform a linear regression (with no intercept) on the known values $\alpha_i$, $i=1,..N$ and the values of $\eta_i$ $i=1,..N$ which minimize $g(R_{X,x},R_{X,y})$ then the optimal solution will also minimize the residuals of this regression, thus:
\begin{equation}
\label{resid}
    h(R_{X,x},R_{X,y})=resid\{\alpha_i\propto\eta_i\}   , i=1,..N
\end{equation}
Having determined the value of $k$ which minimizes \ref{resid}, the gravitational mass, $\mu_X$, can be calculated by rearranging \ref{k} and inserting known parameter $V_S$, as required.

\subsection{Exploring the Oort Cloud}
\label{subsec:Oort-cloud}

With a one au mean distance between Oort comets, the probability for a probe to be at a distance $D < 1$ au from a comet is $(D/1au)^2$, leading to a $1000(D/1au)^2$ probability for 1000 probes. Therefore the probability that at least one probe passes closer than 0.02 au from a comet is 50\%. Such close approach will allow direct imaging of Oort comets, giving their color and detecting outbursts from some of them\citep{Oort} . It will also be possible to explore the self-gravitational dynamics within the inner Oort cloud \citep{Batygin-Nesvorny-2024-a}. And some stellar occultations by Oort comets could potentially be performed, giving their size. There may also be a few interstellar bodies at the outer edge of the Oort cloud and beyond in the direction of Proxima \citep{Neslusan-et-al-2026}.

\subsection{Astrometry}
\label{subsec:astrometric-science}

During its voyage, the probe can observe  target such as the nearby stars (see also Subsection \ref{subsec:astrometry-spectroscopy}).  Compared to the same targets seen from the Solar system, it will see them, and their neighbors on the sky plane, with a different parallax and (for objects viewed by reflection) a different phase angle. One can speak of ``dynamical stereo-astronomy'' since the sky appearance will change along the probe voyage. For astrometry, at the mid term of the voyage, i.e. at $\sim$0.7 pc from Earth, the parallax will be $\sim$6 $\times$ 10$^{4}$ times larger than that seen by Gaia, with a continuous observation possible contrary to Gaia which sees the stars only around 70 times during the mission. At least for bright nearby stars, these observations could  provide a significant improvement in the distance measurements, proper motion and proper motion anomaly of selected targets, the opportunity to detect small mass planets and to measure the orbital inclination of planets known from radial velocity measurements. Among the interesting targets, the planet candidate Gliese 65 b
(GJ 65 b) is orbiting one of the components of the wide binary GJ 65AB (V = 12) at 2.46 pc. While at present  we do not know yet which component GJ 65b orbits \citep{Abuter-et-al-2024-a}, this could be resolved by 
astrometric observations of GJ 65AB by a Proxima mission.

\subsection{Observing Companion Transits en route}
\label{subsec:Transits}

Let us consider a planet orbiting a star with a radius $R_*$ at a distance $D$ on a circular orbit with a semi-major axis $a$. The observer sees, at the appropriate orbital phase, a dip on the stellar light-curve (transit) when the angle $i$ between the binary orbit and the observer is smaller than $\arcsin((R_1 + R_2)/a) $, where $a$ is the separation at the time of eclipse. As the probe is moving on its trajectory, the angle $i$ is varying. If a transit is seen from Earth or from a telescope in orbit, the transit will disappear when $i > \arcsin((R_1 + R_2)/a)$. On the contrary, a transit unseen from Earth or from a telescope in orbit, may appear when the angle $i$ for the probe becomes smaller than $\arcsin((R_1 + R_2)/a)$. Thus, the probe will detect significantly more transits than previously known. In particular, one can predict future transits from Gaia astrometry, combined with radial velocity \citep{2024RSPTA.38230071S}. If favorably located in the sky, the probe could point toward them at the predicted time of occurrence of the transit. 

The geometric probability of transits for a binary system depends on its the orientation of its orbital plane with respect to the probe trajectory. Their orbital period observed by the probe depends on its velocity and on their distance distance $D_p$ from the probe. Their duration depends on the distance $D$, on the companion orbital period $P$, and on the geometry of the binary system orbit as seen from the probe. The impact parameter $b_P$, as seen from the probe, depends on the angle of the probe trajectory with respect to the binary orbital plane and on the probe position on its trajectory.

\begin{figure}[hb!]
\centering
\includegraphics[width=\columnwidth]{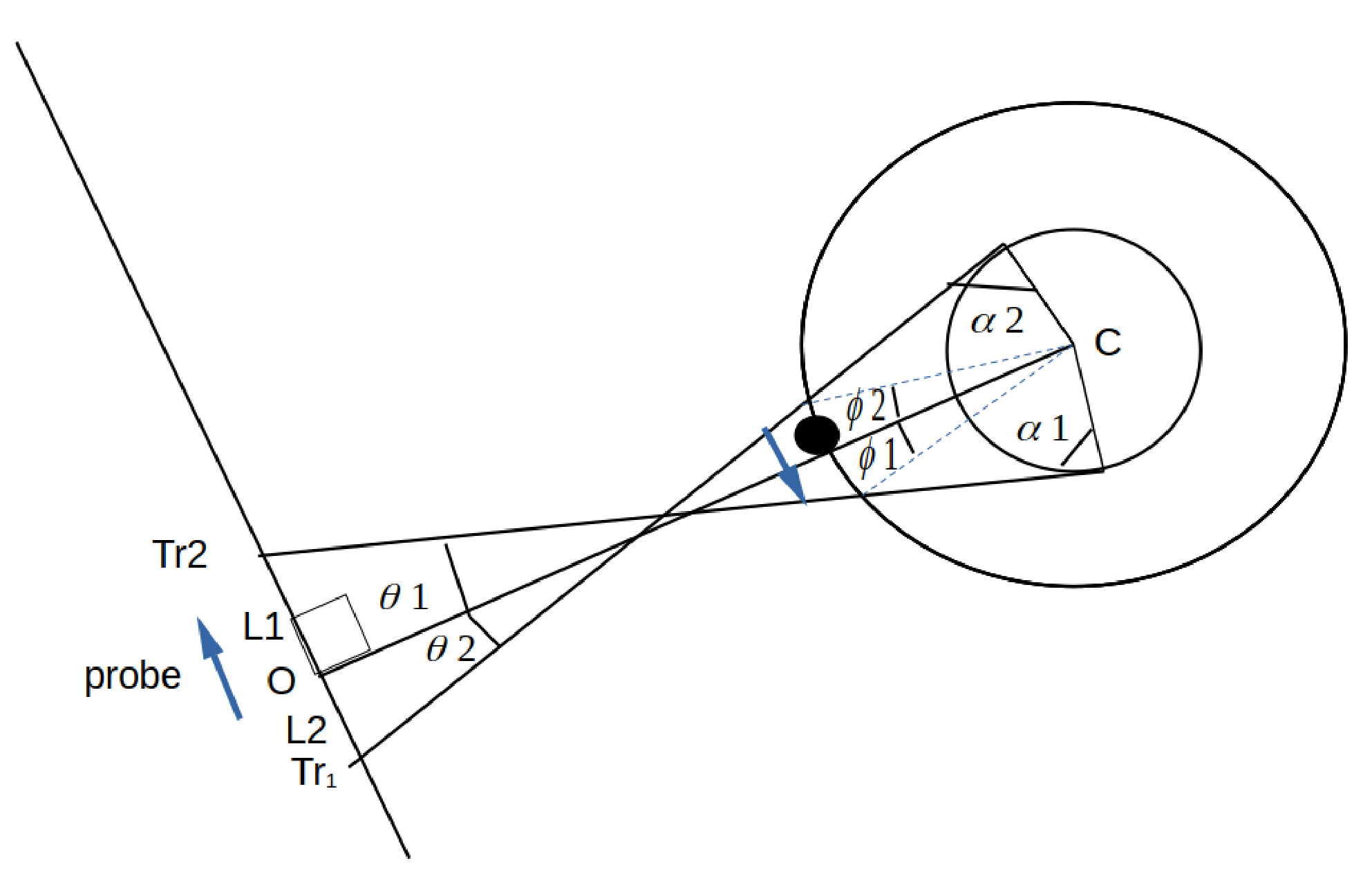}
%%\plotone{Moon_latitude_temp_11.eps}
%%
\caption{ Geometry of the transit seen by the probe in direction antiparallel to the planet trajectory}
\label{fig:antipar}
\end{figure}

\begin{itemize}
    \item Geometric probability of transits
    
    For a binary system at a distance $D$ with an orbit perpendicular to the probe trajectory, the change in $i$ during the 1.4 pc voyage of the probe, the change in the inclination $i$ is $\arcsin 1.4pc/D$. It results then that the geometric probability of observing a transit is multiplied by $1.4pc/D$. There is no change in the geometric probability of transits if the probe trajectory lies in the orbital plane of the binary system.

    \item Orbital period observed by the probe.
    
    Let $P_o$ be the orbital period seen from the Solar system and $D_o$ 
    be the distance of the probe trajectory to the binary system at the point $X_o$ on the trajectory where this distance is at its minimum. 
    
    One has $D_o= D sin{\alpha}$, where $\alpha$ is the angle between the probe velocity and the direction of the binary system seen from the solar system. Then, for a distance $D_{op} (t) =\ O.2c\ t$ of the probe from the solar system at a time $t$ after launch, the distance of the probe to the binary system is $D_p (t) = \sqrt{D^2_o+(D_{op}-D_o)^2}$, where $D_o$ is the distance between the solar system and the point $X_o$. (See Figure \ref{fig:transit-BTS}).
    
\begin{figure}[hb!]
\centering
\includegraphics[width=\columnwidth]{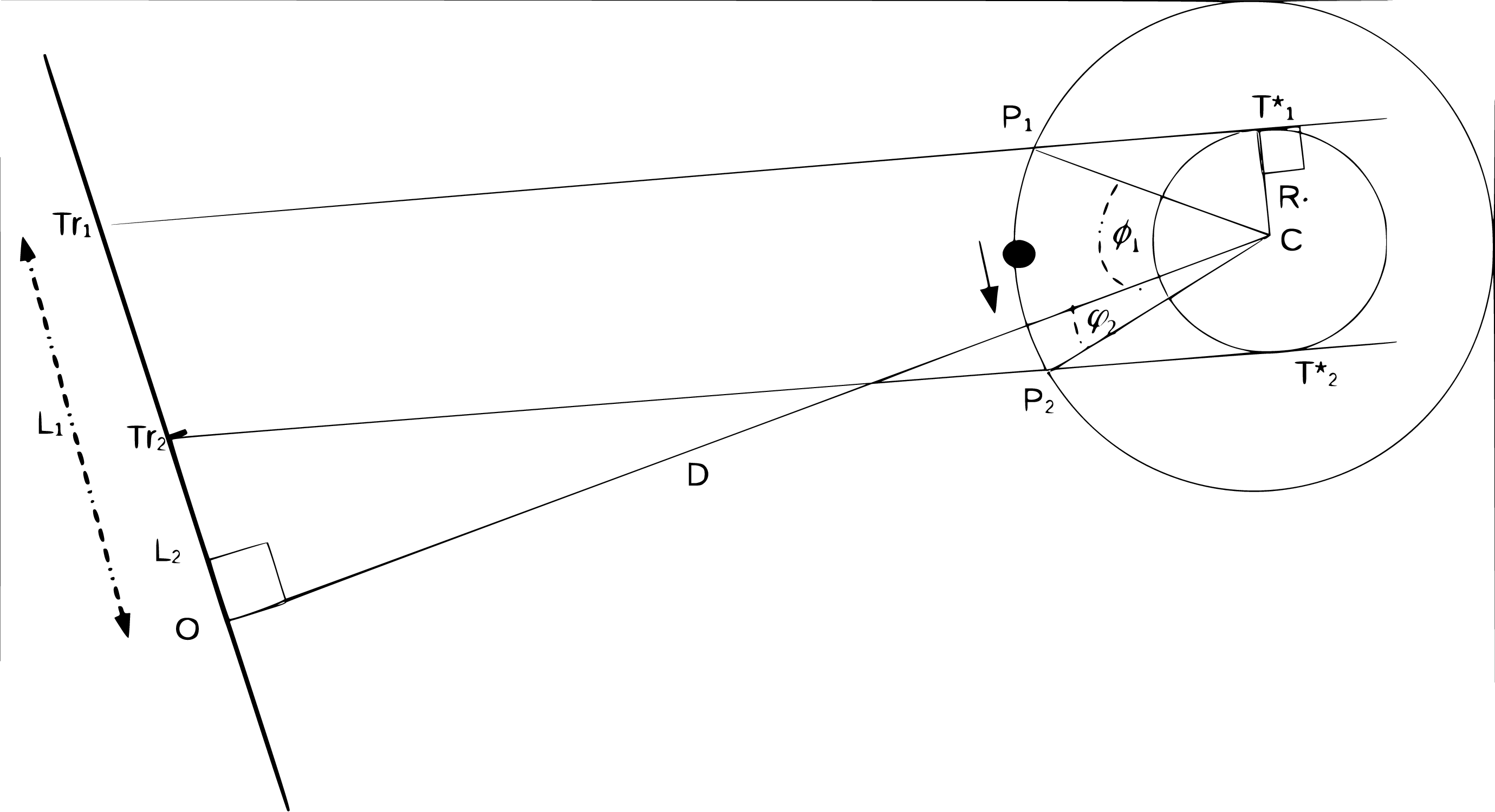}
%%\plotone{Moon_latitude_temp_11.eps}
%%
\caption{Geometry of the transit seen by the probe}
\label{fig:transit-BTS}
\end{figure}
    
    The orbital period seen from the probe is  then $P_p = P_o (1 + (dD_p (t)/dt)1/c)$ .
   
    \item Duration of transits

    Let us consider two extreme cases

\begin{itemize}
        \item The probe trajectory is perpendicular to the binary orbit. Then the probe will see a transit as long as the companion is in front of the observed star, that is a duration $(P/\pi )R_* /a$, the same as for a fixed observer with respect to the star.
        
        \item The probe trajectory lies in the planet orbit. Then the planet and probe trajectories are parallel or anti-parallel. 
   
   \begin{itemize} 
        \item Figure \ref{fig:transit-BTS} shows the geometry of the transit when the planet and probe trajectories are parallel (traveling in the same direction).         
        On the probe trajectory, Tr1 and Tr2 are the points where the transit seen by the probe 
        starts and ends. L1 and L2 are the distance between Tr1 and Tr2 and the point
        O where the distance of the trajectory to the star is minimum. One has
        $L_i =\sqrt{CTr_i^2 - D^2}$
        After some geometry one finds 
        $\ph_i = \arcsin(R_* /a) + \arcsin(L1/D) - \arcsin(R_* /Tr_1C)$
        where the angles $/phi_i$ and the segments $Tr_iC$ are shown on Figure \ref{fig:transit-BTS}.
        Then, the probe sees a transit when the planet goes from points $P_1$ to $P_2$ shown on Figure \ref{fig:transit-BTS}. 
        This takes a time $P(\phi_1 + \phi_2)/(2\pi)$.
        The start and end of the transit are seen after delays $Tr_1P_1/c$ and $Tr_2P_2/c$, where
        $Tr_iP_i =\sqrt{Tr_iC^2 -R_*^2}$.
        The duration of the transit seen by the probe is thus 
         
        $(D_{tr})_{par} =P(\phi_1 + \phi_2)/(2\pi) +  (Tr_2P_12 - Tr_1P_1)c $.

        For a probe of velocity 0, one recovers the standard formula $D_{tr} = (P/\pi )R_*/a $.
         
        \item Case of antiparallel transverse probe velocity and planet velocity. The geometry of the transit seen by the probe is shown on Figure \ref{fig:transit-BTS}. 
        After some geometry one finds for the angles
        $\phi _1$ and $\phi _2$ 
        
        $\phi_i = \pi /2  + \arcsin(L_1/Tr_iC) - \arcsin(R_*/a) - \arcsin(R_*/Tr_iC)$.
        
        Similarly to the case of parallel velocities, the transit duration seen by the probe is then 

        $(D_{tr})_{antipar} = P(\phi 1 + \phi 2)/(2\pi ) $

        From Figure \ref{fig:antipar} one sees that $\pi _i = \alpha _i- \arcsin(R_*/a)$
        and  $\alpha _i = \pi / 2 - \theta _i$

        After some geometry, one has $\theta _i = \arcsin (R_* (1 + L_i /R_*))$.
    \end{itemize}

    Since the transit duration seen from the solar system is $(P/\pi )R_* \sqrt{1-{b_o}^2} /a$ ($b_o$ being the impact parameter  seen from the solar system) and $(P/\pi )R_* \sqrt{1-{b_p}^2} /a$ when seen from the probe, one sees that, by comparing the transit duration seen from Earth and from the swarm, one can infer the sense of the orbital revolution of the planet around its parent star.

\end{itemize}
\end{itemize}

\subsection{Exploring the Interstellar Medium}
During its voyage, in addition to remote observations, the probe will sample the ISM \textit{in situ}.
Analysis of Doppler shifted
absorption lines in the light from nearby solar-type stars reveals the existence of ``local'' clouds in the ISM. In particular, while the solar system appears to be in one cloud, denoted the Local Interstellar Cloud (LIC), $\alpha$ Centauri (and thus presumably Proxima) is in another cloud, the ``G'' cloud \citep{Linksy-et-al-2022-a}. The available data poorly  constrain the size of these clouds and  their interior conditions it is not even clear if our solar system is truly in the LIC. \textit{In situ} observation of the properties of these clouds will be a primary scientific goal for mission science during the long interstellar voyage. 

\subsection{Interstellar Dust}
\label{subsec:interstellar-dust}

The ISM includes dust (small solid particles) in addition to gas.
Although the ISM gas-to-dust mass ratio is thought to about $\sim$100:1, ISM erosion
of near-relativistic spacecraft
is likely to be dominated by dust, with an estimated $\sim$0.5 mm of dust erosion in a near-relativistic voyage to Proxima Centauri \citep{hoang2017interaction,Lubin-2016-a}. For most of the voyage, the probes will travel edge-on (see Section \ref{section:ConOps}), the  armored upper rim  facing forward to be eroded. During the data transmission period after the Proxima flyby, the instrumentation layer will face the solar system and will  be to some degree protected by the sail, and its aerographene backing, which will face forward.
However, dust erosion could be a problem during periods when probes travel face-forward, such as the probes tasked with acquiring approach videos (see Subsection \ref{subsec:approach-imaging}). Although if dust erosion scales linearly with time, yielding a prediction of only $\sim$0.5 $\mu$m of erosion in a 8 day period, dust particles at 0.2 c will tend to punch holes or form craters, roughly the size of their diameter, in structures they impact.

Spacecraft observations of interplanetary dust reveals a population of small particles
with radii as small as 110 nm  \citep{Kruger-et-al-2019-a}. A columnar gas density estimate of 3 $\times$ 10$^{21}$ protons m$^{-2}$ all the way to Proxima \citep{hoang2017interaction} corresponds to a dust columnar number density estimate of 2.5 $\times$ 10$^{9}$  m$^{-2}$ 110-nm dust particles in a 20 year voyage, causing holes or craters in $\sim$10$^{-4}$ of the area of the forward facing rim in that period, and roughly $\sim$10$^{-7}$ of the area of the probes used in the 8-day approach video production.

 Recording the impact rate due to collisions with ISM ``dust'' (here, any particles bigger than single molecule) or other hazards will both inform this mission's operation and also provide the first high-fidelity map of the dust density all the way between Sol and Proxima, thus providing much basic  astronomical and stellar cartographic intelligence to the scientific community and the world at large long before arrival. With the above dust models, probes traveling edge on, with an exposed rim area of 0.08 m$^{2}$, should experience a dust impact on average every 3.2 seconds with an energy of $\sim$0.04 J.
 This energy should be enough to count impacts with acoustic sensors in the rim structure, enabling the determination both the number density and mass distribution of dust particles in a wide range of size and mass  from every probe in the swarm.
 
\section{Fundamental Physics}\label{Sec:Fundamental_Physics}

\subsection{Tests of Gravity}
The analysis in \citep{Banik-Kroupa-2019-a} shows that in the MOND theory of alternative gravity the orbit of Proxima Centauri about the primary Alpha Centauri system would be perturbed, with the integrated orbital motion over a decade being $\sim$5000 km (for MOND) and $\sim$3600 km (for Newton), with the difference being $\sim$1600 km $\times$ (t/1 decade)$^{2}$. If the probes can determine their position to within 1 km (say, by observing X-ray or radio pulsars, and also observing the planet's position at the arc-second level), then those relative accelerations could be possibly determined by 4 swarms of probes separated by 1 year each, looking for a total quadratic position change difference of 250 km.

\subsection{Interstellar Gravitational Wave Detection}
\label{subsec:grav-wave-detection}

Gravitational radiation in the cosmos is thought to be dominated by the inspiral and merger of binary systems with ultra-compact objects; in multi-year periods these are thought to be dominated by radiation from SuperMassive Black Hole Binaries (SMBHB) 
\citep{Lommen-2015-a}. Sufficiently small SMBHB orbits will inspiral due to orbital decay caused by the loss of angular momentum through emission of gravitational radiation.
During the binary inspiral process the gravitational wave (GW) frequency steadily increases (causing GW ``chirps'') up to the orbital frequency of merger, f$_{merger}$, which is also the time of the largest gravitational wave strain from the merger.

\begin{figure}[hb!]
\centering
\includegraphics[width=\columnwidth]{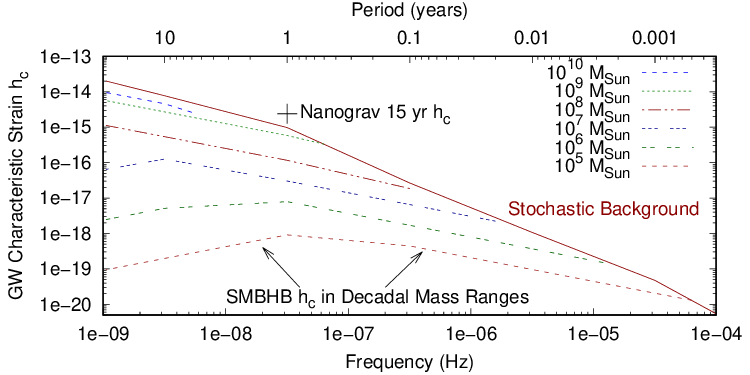}
%%\plotone{Moon_latitude_temp_11.eps}
%%
\caption{Predicted gravitational wave characteristic strains based on statistical models of the SMBHB number density \citep{Pesce-et-al-2022-a}, here broken into decadal mass ranges. At a given mass, the number of sources declines going along the curve to the right (i.e., towards higher frequencies); 
each curve stops at the frequency step  where the expected number of sources in that mass decade falls below 1. These models agree reasonably well with the characteristic strain estimate, h$_{c}$, from the 15 year Nanograv pulsar timing analysis, shown here scaled to a nominal 1 year period \citep{Agazie-et-al-2023a}. 
\label{fig:grav_wave_mass_integration_24}
}
\end{figure}

Figure \ref{fig:grav_wave_mass_integration_24} shows models of the expected amplitude of the gravitational wave signal on the Earth-Proxima baseline distance. 
These models are in rough agreement with recent determinations of the stochastic gravitational wave background by pulsar timing arrays (PTA), which indicate that
this signal will be dominated by  binary
masses $\gtrsim$ 10$^{8}$ M$_{\bigodot}$ and orbital periods of order a few years to a few decades
\citep{Sato-Polito-et-al-2024-a}. In the 1 to 10 year period range, the dimensionless characteristic GW strains should be of order 10$^{-15}$ to 10$^{-14}$.

The gravitational wave background is expected to have a strong red-noise spectrum, implying that interstellar ranging over baselines of many ly will mostly be sensitive to gravitational waves with periods comparable to the baseline length, and thus to the merger of SMBHB. 
The swarm-Earth baseline lengths will thus be comparable to the GW wavelengths for the strong likely GW signals; in this ``short baseline'' regime the total baseline length change due to a GW will have amplitudes of the  baseline length times the GW strain amplitude.
The Earth-Proxima baseline is $\sim$4 $\times$ 10$^{16}$ m;
the gravitational wave flux estimates in Figure \ref{fig:grav_wave_mass_integration_24} 
suggest that this baseline (and the Earth-swarm baselines) could change by 100s of meters due to GW over multi-year periods.

It is not likely that these relatively small multi-year range variations could be observed directly by a single interstellar swarm, as this signal would have to be separated from range changes due to variations in the ISM drag, but it should be possible to create normal points for the Solar System-Proxima system range from each  passage of swarms through the Proxima system. As  multiple swarms are sent through the Proxima system over time, detection of the gravitational wave strain should become possible. 

\section{Arrival Science}
\label{sec:arrival-science}

\begin{table*}[ht]
\centering
\begin{tabular}{ |c|c|c|c| }
\hline
Event & \multicolumn{1}{|c|}{Distance}  & Time from & Assumed \\
      & \multicolumn{1}{|c|}{}        & Encounter    & Abs. Mag. H    \\
\hline
%%\multicolumn{5}{|c|}{ }\\
%%\multicolumn{5}{|c|}{ }\\
\hline
Detect  Proxima~b       &  $\sim$10,000 au & $\sim$300 days        & -4.0    \\
Detect Proxima~d        &  $\sim$6000 au   & $\sim$170 days        & -1.6    \\
Detect Proxima~c        &  $\sim$1400 au   & $\sim$40 days         & -7.0    \\
Resolve Proxima disk    &  $\sim$271 au     &       7.8 days  &  - \\
Detect 100 km Asteroid  &  \multirow{2}{*}{$\sim$27 au} &  \multirow{2}{*}{18.5 hr}  &  \multirow{2}{*}{9.0}\\
in Prox. b orbit & & & \\
Resolve Proxima~b disk  &  $\sim$21  au     &  14.6 hr  &   -4.0\\
\hline
\end{tabular}
\caption{Imaging possibilities before or after the Proxima encounter.
The search for other planets in the system would be possible in the months around encounter,  and in the few days around encounter an encounter video could be obtained monitoring atmospheric changes on all 3 planets. (Note that Proxima c, if it exists, is sufficiently far from its star to make it relatively dim, even though it would presumably be considerably larger than Proxima b.)}
\label{table:approach-and-departure}
\end{table*}

\subsection{Approach Imaging}
\label{subsec:approach-imaging}

Roughly 8 days before the encounter, at a distance of $\sim$420 au, the primary probe imaging apertures would be able to resolve the Proxima b Hill sphere (which has a radius of $\sim$150,000 km) and a search for Proxima b moons can begin. 
This would begin the approach imaging period and initiate detailed observations of the Proxima system. 

As the approach monitoring probes will have to change their attitude from edge-on to face-forward during this period (see Section \ref{section:ConOps}) they will suffer more drag and will fall behind the main swarm during the approach period. As this drag separation should only be order 10$^{5}$ km, and as the main communications apertures of these probes will be facing the main swarm, transfer of approach data to the main swarm should proceed continually during this period. 

At a distance of $\sim$21
au, 14.6 hours before encounter, Proxima b will be resolved and it would be possible to
begin a planetary approach video for it and the other planets. 
It should also be possible in this period to observe, with forward scattering, the dust rings detected by ALMA around Proxima \citep{Anglada-et-al-2017-a}. Probes dedicated to approach imagery should be equipped with both narrow angle and wide angle cameras to take advantage of these opportunities. 

Proxima itself is an active flare star, and, as described in Subsection \ref{subsec:proxima-science}, at roughly 8 days before encounter it will also be possible to resolve the star's disk, locate flares on its surface, and begin acquiring a stellar approach video. (The monitoring of flare activity could begin much earlier, but will be limited by the need to minimize exposure of optical instruments to ISM wind erosion.) Given the relative brightness of Proxima compared to its planets,  certain probes should be designated for approach imagery and given a subset of apertures with suitable optics to observe Proxima directly.

As discussed in Subsection \ref{subsec:dynamic-targeting}, DT needs lookahead data to reconfigure the upcoming observing schedule. In near-Earth satellite observing lookahead data can be provided by special, lower-resolution sensors, or acquired from stand-off sensors (e.g., from geostationary sensors for instruments in low Earth orbit \citep{Kangaslahti-et-al-2026}). 
With spacecraft swarm missions the  swarm can instead reconfigure itself to provide lookahead data, but the Proxima swarm would have very little time (a matter of seconds) to adjust observing schedules during the close approach to Proxima b itself.

 Instead, imaging and other data from the approach period can be used as DT lookahead data. DT schedule reconfigurations will include both targeting adjustments based on improvements in the Proxima planetary system ephemeris, the imaging of new object candidates using results from the search for previously undiscovered objects, and imaging changes based on an initial characterization of planetary conditions, such as the discovery of cloud banks or surface features. DT will also be applied to the approach period observing schedule itself, primarily to focus on newly discovered features and, if necessary, to replace damaged forward-facing imaging probes. 

%
%% new image
%%
\begin{figure}[hb]
\includegraphics[width=\columnwidth]{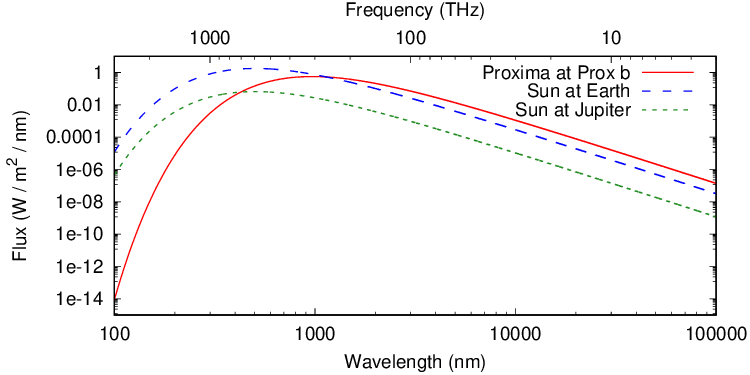}
\caption{The illumination (proxlight) from Proxima at Proxima~b during non-flare periods compared with the solar illumination in the solar system (this and subsequent images ignore any spectral absorption or transmission lines in the stellar spectra). Very approximately, the visual-band proxlight at Proxima~b, while  
54 times weaker than sunlight at Earth (see Table \ref{table:Centauri_Planets}), is comparable to sunlight at Jupiter, while in the red and IR the proxlight at Proxima~b is brighter than sunlight  for wavelengths $\gtrsim$  427 nm (at Jupiter) and $\gtrsim$ 1143 nm (at Earth).}
\label{fig:Proxima-Earth-illumination}
\end{figure}

\subsection{GigaPixel Imaging in the Proxima System}
\label{subsec:GigaPixel-Imaging}

%%\subsection{Near-Light-Speed Imaging}
%%\label{subsec:relativistic-imaging}

A single 200-mm aperture observing at optical wavelengths has a diffraction limited resolution of order 0.4 arc-seconds, providing a potential for gigapixel imagery in the Proxima system, which would revolutionize the study of the Proxima planets. A swarm with a few hundred surviving members and a total extent of 10$^{5}$ km should have some members passing within $\sim$10$^{4}$ km of the target planet (roughly its diameter), providing a potential Proxima b resolution  of $\sim$20 meters. Smaller sub-swarms directed at other planets (Subsection \ref{section:ConOps}) may not make such close approaches, especially for Proxima d, which is likely to have larger ephemeris errors than Proxima d. We assume these secondary targets will have miss  distances up to 10$^{5}$ km, providing resolutions of $\sim$200 meters. 

Table \ref{table:Centauri_Planets} and Figure \ref{fig:Proxima-Earth-illumination} show that the ``proxlight'' at Proxima d is roughly comparable to the  light of the Sun at Earth, while Proxima b at its mean distance would have proxlight  $\sim$54 times weaker than terrestrial sunlight, and for Proxima c at its nominal distance the proxlight would be roughly 10 times brighter than the light of a Full Moon on Earth. 
The close approach imaging problem can thus be compared to that faced by the recent Artemis II mission\footnote{\url{https://www.nasa.gov/artemis-ii-multimedia/\#images}},  except that these probes will be traveling much faster. 

Artemis II imagery taken with Nikon D5 cameras with aperture areas comparable to the Coracle cameras, used International Organization of Standards (ISO) Recommended Exposure Index (REI) speed ratings ranging from 400 (for sunlit shots with exposures of order 1 ms) to 51,200 (for the ``Hello World'' photograph, Art002e000192, taken with a 250 ms exposure of the night-side Earth lit by the full Moon). Allowing for the differences in illumination, comparable Proxima d images could be done with 1 ms exposures, Proxima d images with 54 ms exposures, both with REI of 400, and Proxima c images could be done with 25 ms exposures with a ISO REI of $\sim$ 50,000 (for an ISO speed of 400 the exposure would need to be longer than the entire duration of the close approach period).  The maximum resolution of imagery from a light-sail swarm would thus be substantially better than is possible from  optical telescopes in the solar system observing from a distance of 4.24 ly \citep{Turyshev-2026-a}.

At 0.2 c  a single camera aperture would be  able to acquire order 2 - 4 full exposures  at close approach for Proxima b or c, and order 100 exposures at Proxima d. 
As the probes will move $\sim$60 m per $\mu$s, to avoid image smearing and 
obtain high-resolution color images and spectra these exposures will have to be assembled from sub-exposures with durations of order one micro second, requiring
extensive  use of TDI and VSI to correct for perspective and redshift changes during the full exposure  (see Subsection \ref{subsec:TDI-VSI}). This process, spread over hundreds of probes, will produce a huge amount of data, as much as a terabyte integrated over the entire swarm. Clearly the computations required to generate full-exposure images will have to be performed by the swarm at Proxima Centauri; this will reduce the amount of imaging data that has to be returned to Earth by roughly 4 orders of magnitude.

%
%% new image
%%
\begin{figure}[hb]
\includegraphics[width=\columnwidth]{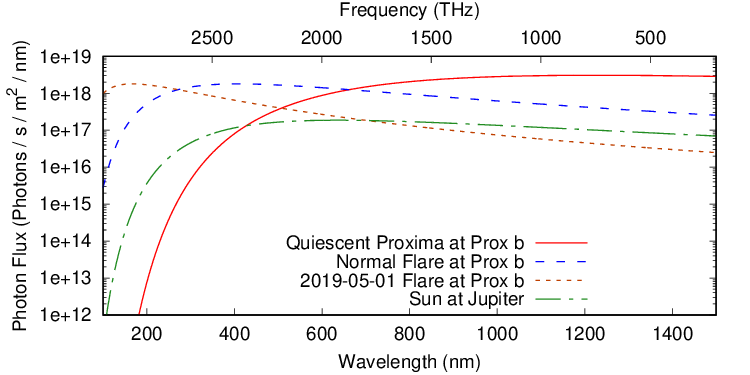}
\caption{Proxima has $\sim$ daily optical flares, with brightness temperatures of order 9000 K, and 
occasionally much stronger (and hotter) flares, such as 
the flare observed on May 1, 2019
\citep{MacGregor-et-al-2021-a}. These flares substantially increase the blue and UV light available at Proxima~b; there is a decent chance of a bright optical flare during the 0.8 day period while Proxima~b is resolved by the Coracle cameras.} 
\label{fig:Proxima-flare-light}
\end{figure}
\begin{figure}[hb]
\includegraphics[width=\columnwidth]{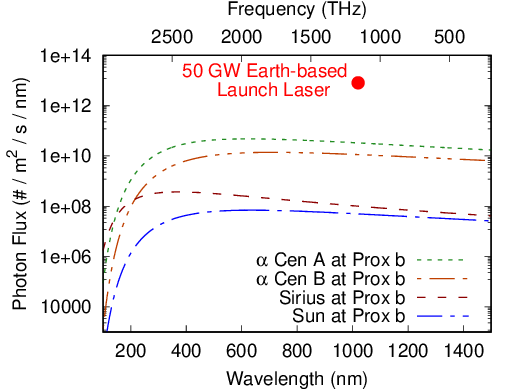}
%%\caption*{Bright point sources at Proxima Centauri}
\caption{Sources suitable for transmission spectroscopy of Proxima b. 	$\alpha$ Centauri A and B are the only natural sources in the Proxima b sky  bright enough in the visual band to support 10-meter resolution sampling of its atmosphere. Sirius and other bright stars could also be used for this purpose, but with larger integration times and thus worse resolution. 
The terrestrial launch laser, if it is turned on for the encounter, can be used as  an ``interstellar flashlight,'' illuminating Proxima b and other targets along swarm's trajectory (Subsection \ref{subsec:night-side-imagery}) and also would also be an excellent source for transmission photometry  (Subsection \ref{subsec:transmission-spectroscopy}).  }
\label{fig:transmission-spect-1}
\end{figure}
%%%%%

%%
%%
\begin{figure}[hb]
\includegraphics[width=\columnwidth]{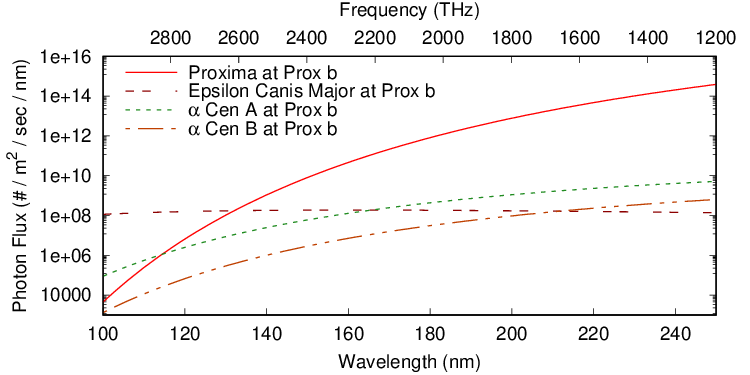}
%%\caption*{Bright point sources at Proxima Centauri}
\caption{EUV Sources suitable for transmission and reflection spectroscopy of Proxima and Proxima b. 	The  EUV source Epsilon Canis Majoris ($\epsilon$ CMa) is normally the brightest source in the sky at Proxima b, substantially brighter than the (non-flare) EUV emission from Proxima itself. 
It would thus be an ideal source for EUV transmission spectroscopy of the outer layers of the atmospheres of both of these bodies.}
\label{fig:transmission-spect-2}
\end{figure}
%%%%%

\subsection{Night-Side Imagery}
\label{subsec:night-side-imagery}

Assuming that there are no large undiscovered moons orbiting Proxima b, the drive laser and AB $\alpha$ /Centauri will be the brightest light sources in the night-time Proxima b sky, and will  thus be crucial to imaging the Proxima b night-side. As Proxima b is likely to be tidally locked, and the night-side could thus remain  dark for all future missions, probes, trajectories and  mission timing should all be developed to provide the best possible low-light and IR imagery of the planet's night-side.

Figure \ref{fig:transmission-spect-1} shows that the drive laser in the solar system can be a significant light source at Proxima Centauri, if it is available for encounter support. Here we use conservative estimates
from \cite{Parkin-2018-a}, with a laser optical power of 50 GW at 1064 nm and a laser beamwidth of 0.5 mas, equivalent to a transverse spot size of $\sim$100,000 km in the Proxima system. This yields a monochromatic flux the laser beam of $\sim$1.6 $\times$ 10$^{-6}$ W m$^{-2}$, or 
$\sim$8.1  $\times$ 10$^{12}$ photons m$^{-2}$ s$^{-1}$,
about 2 $\times$ 10$^{-3}$ of the brightness of the full Moon on Earth. 

$\alpha$ Centauri A and B combined provide 
$\sim$(4 or 5) $\times$ 10$^{10}$ photons m$^{-2}$ s$^{-1}$ nm$^{-1}$ at 1000 nm and 500 nm, respectively. These are of course broad-band sources; integrating over a band-width of 100 nanometers $\alpha$ Cen can provide a photon flux comparable to that from the drive laser. Even if the drive laser and $\alpha$ Cen are combined they are unfortunately too dim to allow for full resolution images in the time available at close approach; images will have  to  have reduced angular resolution to obtain an adequate SNR.

Assuming that the surface of Proxima b is a Lambert scatterer with an albedo, $\alpha$, of 0.1, then a 100  $\times$ 100 km surface element at a camera  distance of 0.01 au would return about 80 photons m$^{-2}$ s$^{-1}$ from the drive laser, and about 50 photons m$^{-2}$ s$^{-1}$ from  AB Centauri (assuming a 100 nm spectral bandwidth at 500 nm). As 0.01 au is traversed in 25 seconds, a single imaging element would have a very marginal SNR with a 100 km resolution, but a collection of apertures observing together could gain a reasonable SNR, and with the AB Centauri light color images would be possible. We thus regard 100 km as a reasonable estimate for the resolution limit of night-side imaging (although this could be exceeded if part or all of the night-side has a high albedo or if another light source is available). 

\subsection{Bistatic Laser Ranging}
\label{subsec:bistatic-lidar}

The drive laser signal should be pulsed or modulated to allow for bistatic laser ranging (lidar) of objects in the Proxima system, something that cannot be done with starlight. As the laser spot size will be comparable to the errors in the Proxima b ephemeris when it is sent from the solar system 4 years before the first flyby, it may be necessary to scan the laser spot across the likely locations of Proxima b (and, also, of the swarm itself).  The swarm will be easily able to receive laser pulses (or other modulation) from the drive laser when it is in the laser spot even with relatively small laser receivers, enabling it to determine the clock offset between its clock and solar system barycentric time. By modulating the laser beam with an identifier for the spot position, and recording that together with the  the direction and time of arrival of laser light reflected from the  target, it should be possible to determine both which spot position the target is in, and thus its transverse position to within about 10$^{5}$ km, and also the relative time delay between the direct and reflected arrival times, and thus the bistatic delay difference to the target, possibly to within a nanosecond level. This technique could greatly improve the ephemerides of objects orbiting Proxima, and  would be so scientifically valuable that if necessary a selection of probes should be rotated before the Proxima encounter to have their instrument-side facing in suitable directions to acquire it.

\subsection{Transmission Spectroscopy}
\label{subsec:transmission-spectroscopy}

Transmission spectroscopy can be done at Proxima b using natural and artificial sources, and will,  through the search for spectral lines of biomarkers and technosignatures, likely provide the best means of establishing the existence of a biology or even a technological society on Proxima b. 
The swarm should be numerous enough to provide multiple observations of transits  through the Proxima b atmosphere, enabling  determinations of its global density structure and detection of any cloud layers. 

$\alpha$ Cen A and B will be a bright close double in the Proxima b sky, with visual magnitudes of $-6.6$ and $-5.3$, respectively. Figure \ref{fig:transmission-spect-1} shows that these will be the best natural sources for transmission spectroscopy of the Proxima b atmosphere.
 These two sources, with apparent diameters of 9 and 6 meters at a distance of 10,000 km from Proxima b, 
would enable 10-meter resolution sampling of two close raypaths in the Proxima~b atmosphere. 
To avoid motion smearing, such transits must be sampled with $\sim$200 nanosecond resolution, requiring  $\gtrsim$10$^{9}$ photons m$^{-2}$ nm$^{-2}$ s$^{-1}$ 
with an $\sim$0.01 m$^{2}$ effective aperture. AB $\alpha$ Cen are the only stars in the Proxima sky bright enough to provide a reasonable SNR for such transits in the visible and near-IR spectral regions.
With A and B
orbital separations between  11.2 to 35.6 au, or 9 to 28 km at a distance of 10,000 km, each transit would provide that density structure at two close points in the atmosphere, enabling horizontal gradients to be determined. 

The terrestrial drive laser would also be very bright (see Figure \ref{fig:transmission-spect-1}, and would make an excellent source for transmission photometry. The drive laser can in addition be used as an ``interstellar flashlight,'' illuminating targets along swarm's trajectory (see Subsection \ref{subsec:night-side-imagery}).

%% 1e13 photons / m
%%  flash_light_beam  = 2.5000e-09 radians                  =     0.5157 mas 
%% flash_energy_flux  = 1.5806e-06 W / m^2 
%% flash_photon_flux  = 8.1162e+12 # / m^2 / s 
%% The heat provided by the full Moon is negligible, estimated at about 3.1 milliwatts per square meter.

\subsection{Proxima Helioscience}
\label{subsec:proxima-science}

Proxima is a frequent flare star, with one or more flares per day $\gtrsim$ 5\% of its normal luminosity \citep{MacGregor-et-al-2021-a} and occasional ``superflares'' brighter than its normal luminosity \citep{Howard-et-al-2018-a}. 
Observation of the location of these flares relative to the stellar disk would be possible up to 8 days before and after encounter (see Table \ref{table:approach-and-departure}); the swarm should thus be able to provide valuable observations of the details of possibly a few dozen moderate intensity flares. After the flyby, when the swarm is communicating with Earth, it will be better able to observe Proxima, and can provide observations of the Proxima flares (which may be beamed) from a totally different vantage point than monitoring from Earth. 

Figure \ref{fig:Proxima-flare-light} shows that 
Proxima  flares are much hotter (and thus bluer) than the quiescent star itself. These flares  can thus used as ``interstellar flashbulbs,'' providing light for blue and UltraViolet (UV) imaging of the planet's day-side in spectral bands where Proxima is otherwise a very dim source. Unfortunately, even though there are daily flares it is unlikely this light would be available during the few minutes of the actual closest approach, and so the best UV light would likely occur when the swarm was multiple au from the planet, yielding resolutions no better than continental-scale. However, if the timing of some flares is connected to the planet's orbit (e.g., if there is a flux tube connecting Proxima and Proxima-b) it may be possible to provide high quality UV imaging through an appropriate scheduling of the time of the closest approach. In that case, it may also be possible to arrange for probes to fly through the flux tube, and thus observe it \textit{in situ}.

As can be seen in Figures \ref{fig:transmission-spect-2} and  \ref{fig:Proxima-flare-light} Proxima itself is very dim in the Extreme UV (EUV) except during flares. The swarm can thus do EUV transmission spectroscopy of the star itself   using the very bright stellar EUV source $\epsilon$ CMa, which is brighter at Proxima~b  than the quiescent  Proxima Centauri itself at wavelengths $\lesssim$133 nm. This would provide a novel means of determining the density and temperature structure of the upper layers of this star. 

%%\subsection{Interstellar Flashbulbs}
\subsection{Impact Spectroscopy} 
The distance between probes at the fleet's center could be as little as a few thousand kilometers, and it is possible that one or more Coracle probes would enter the Proxima-b  atmosphere, if it exists, or impact the surface, if it does not.   Each impact would be a different type of interstellar flashbulb, releasing $\sim$6.5 terajoules or $\sim$1.5 kilotons of TNT equivalent. (Note that the Earth experiences blasts with this kinetic energy  in its upper atmosphere on a regular basis with no ill effects.)  The heavier elements in the probe can be expected to release hundreds of MeV per colliding heavy atom, which is energetic enough to excite or even disrupt atomic nuclei.   The resulting flash, which could be predicted in advance by the swarm, would be potentially observable from the visible spectrum into the Gamma Ray bands by nearly the entire fleet, and would yield important spectroscopic data about Proxima b's composition.

\subsection{Proxima b geocorona}
The Earth's geocorona (or exosphere), the planet's outermost neutral atmosphere, stretches well beyond the lunar orbit \citep{Bertaux-et-al-1995-a} and is composed mostly of  H and He atoms \citep{Kameda-et-al-2017-a}. 
This is visible primarily via far-ultraviolet light (Lyman-alpha) from the Sun that is scattered from neutral hydrogen. It will be hard to observe any geocorona around Proxima b by light from the quiescent Proxima b, but 
Proxima probes could detect these atoms \textit{in situ} and also observe them by Proxima flare light. 

\subsection{Bio- and Technosignatures} 
With the planned number and distribution of probes, it should be possible to have at least one probe pass within one diameter of the target planet, allowing for both detailed imaging of Proxima~b and also transmission spectroscopy of that planet's atmosphere. This will enable an intensive search for bio- and technosignatures there
\citep{Lingam-Loeb-2021-a,Schwieterman-et-al-2024-a,Eubanks-et-al-2024-b,Eubanks-et-al-2024-d}.

Reflected light imaging can be used directly to search for the vegetation red edge (VRE), the higher reflectivity of land vegetation in the visual red and the near-IR \citep{Burr-et-al-2026-a}, which is thought to increase over geological time \citep{OMalley-Kaltenegger-2018-a}, and therefore could possibly be used, if detected, to constrain the age of the biosystem. Although the VRE is generally thought to be a feature of land-based vegetation, it may also be found with floating vegetation \citep{Murakami-et-al-2025-a}. The VRE should be an easy target for Proxima-b swarm imaging; it should be possible to map it in detail on the day-side of the planet, and also to map it coarsely on the night-side (see Subsection \ref{subsec:night-side-imagery}).

The day-side surface resolution of $\sim$60 m would be $\sim$4 orders of magnitude better than albedo maps from future Earth telescopes \citep{Berdyugina-Kuhn-2019-a}, enabling detailed spectroscopic biosignature searches.
Additional biosignature candidates include:    
\begin{enumerate}
    \item  The IR step biosignature from chlorophyll  and the UV step technosignature from solar cells \citep{Lingam-Loeb-2017-a,Kopparapu-et-al-2024-a} should be mappable at high resolution. 
    \item  Biologically driven atmospheric lines such as 

- H$_{2}$O (from 700 to 2000 nm), CO$_{2}$  (from 1400 to 2100 nm), O$_{2}$ (lines at 760 and 1260 nm), CH$_{4}$ at 1650 nm.

- Dimethyl Sulfide (DMS), a by-product of microbial surface activity \citep{MPC21}.

- Color features caused by biopigments of UV-resistant atmospheric microorganisms \citep{Coelho}.

-  Technosignature lines, including  those due to chlorofluorocarbons, NF$_{3}$  or SF$_{6}$ \citep{Schwieterman-et-al-2024-b}, which should all be detectable with both reflection and transmission spectroscopy. 

These atmospheric and surface biosignatures and technosignatures are by themselves only candidates since there can be abiotic false positives.

\item With the high resolution imaging possible at closest approach large biological features such as coral reefs, or large technological features such as transportation networks, or night-time city lights, could be unambiguously imaged.
\end{enumerate}

\subsection{Radio Science}

Relatively low frequency radio waves are produced in planetary systems by the interaction of charged particles from  the stellar wind with the planetary and stellar 
magnetic fields.
Such radio emissions have been detected from Proxima \citep{Perez-Torres-et-al-2021-a}, driven by the Cyclotron Maser Instability (CMI) \citep{Treumann-2006-a}. 
The total Proxima radio emission is of order $10^{12}$ W between 200 MHz and 1.6 GHz, peaking at 1.6 as would be expected from  CMI at the cyclotron
frequency for Proxima's predicted magnetic field intensity of $\sim$0.06 Tesla, and apparently varying in amplitude with the orbital period of  Proxima b. This radio flux should be easily detectable by a small on-board antennas, and its generation could also be observed \textit{in situ}, especially if there are flux tubes (analogous to Jupiter-Io flux tubes) between Proxima and Proxima b or d. 
Measurements of 
the intensity and frequency of CMI radio emissions will provide information about the stellar and planetary magnetic fields in the Proxima system \citep{Zarka-et-al-2001-a}, complementing direct measurements from magnetometers carried by the swarm.

The Earth has a very strong Auroral Kilometric Radiation (AKR), emitting 1 to 10 MW at frequencies between 50 and 500 kHz
\citep{Mutel-et-al-2003-a}. If Proxima b is a magnetized body then it very likely also has a CMI driven AKR; these  emissions could easily be detected by a \textit{in situ} swarm with a suitable radio receiver as it passes by the planet. CMI-driven radio emissions should be common with magnetized exoplanets \citep{Vanhamaki-2011-a}, and there could be multiple AKR type sources in the Proxima system. 

It has been shown that Proxima's stellar activity could  trigger CMI-type radio emission from  Proxima b itself \citep{Pena-Monino-et-al-2024-a,Pena-Monino-et-al-2025-a}, which should straightforward to measure during close approaches to the planet. These radio emissions could also be correlated with aurora on that planet, which could be observed directly in the visual, UV or EUV bands. It would also be important to observe the planet's magnetic field directly \textit{in situ}; 
if the radius of the magnetopause is smaller than the planet radius, then
there could be direct precipitation of damaging particles onto the planetary atmosphere, which could hamper its  
habitability.

\subsection{Small Bodies}
Orbital analysis indicates that asteroids in 1:1 mean motion resonances (Trojan type asteroids) with the planet Proxima c may have stable orbits and thus there could be a Trojan asteroid belt in that planet's orbit \citep{Ipatov-2023-a}.
Proxima c may also have a ring system, leading it to be anomalously bright in the IR
\citep{Gratton-et-al-2020-a}. Beginning approximately 1 day before the Proxima b encounter, imaging of the Proxima system can be used 
to search for exterior planets, asteroid belts, and planetary moons and ring systems using the forward-facing approach imaging probes; this observational search can be continued with additional through the entire encounter period.

\section{Operations beyond Proxima}
\label{sec:ops-beyond-Proxima}

After the encounter with the Proxima system the swarm will orient itself to travel face-down to communicate with Earth. An immediate goal  will thus be to evaluate and select data for return  to Earth (see Section \ref{Sec:data-selection}), while the swarm will also need to  move the swarm members to account for  the dispersion or loss of probes in the Proxima system.  With the probes traveling face-down, the swarm will also be well positioned to observe the receding Proxima system, acquire a departure video, and continue the search for new objects in the Proxima system. After the Proxima encounter, the probes can of course observe the interaction of the Proxima heliosphere and the ISM on the far side of the system, and then conditions in the further ISM itself, for  as long as communications can be maintained. 

Roughly 1 year after the Proxima encounter  the swarm will have a distant encounter with the $\alpha$ Cen AB system, with a closest approach of $\sim$10,000 au. At closest approach, planets in the HZ of either Cen A or Cen B  will have angular separations of $\sim$20 arc-seconds per au from their primary, and magnitudes (for an Earth-like planet in the HZ) of $\sim$16, and so could be observable by the Proxima swarm, which could also monitor the A and B stars for starspots, stellar flares, and  stellar transits by objects within that system. If there is a companion expedition to $\alpha$ Cen timed to arrive 1 year after the Proxima expedition, coordinated simultaneous observations from the Proxima expedition (and from Earth) could enhance the scientific return from that system. 

The star Proxima would only be able to deflect the probe trajectories
by $\sim$5 arc minutes; there are no stellar systems with 100 pc of Proxima that are that aligned with the Sun-Proxima trajectory, and thus no other systems 0.2 c probes could realistically reach with a gravity assist from Proxima. Unless nomadic free-floating planets can be found almost directly behind Proxima, further interstellar exploration will require separately launched missions.

\section{Discussion}
\label{Sec:Discussion}

In a recent review, \cite{Turyshev-2026-a} discussed the difficulty of obtaining even $\mu$as resolution images of exoplanets from solar system observations, equivalent to a transverse resolution of $\sim$200 km at Proxima Centauri. He concludes that no feasible solar system technique could produce even 100 pixel images of a nearby exoplanet, while the solar gravitational lenses could produce megapixel images. 

Only \textit{in situ} exploration has the potential to provide gigapixel images of nearby interstellar targets with technology likely to be available in the next 75 years. Direct exploration also offers many other opportunities to study the target planet, Proxima b, the Proxima system in general, and the interstellar environment between the solar and Proxima systems.

 \cite{Turyshev-2026-a}  also discusses difficulties in producing high resolution images with \textit{in situ} imaging by lightsails, these being
\begin{itemize}
    \item 
    \textit{Encounters at 0.2 c provide only a very short time for observations.}  
    With high dynamic range imaging and TDI and VSI (as discussed in Subsection \ref{subsec:TDI-VSI} and \ref{subsec:GigaPixel-Imaging}), composited images can be created minimizing image smear for the needed exposure times.
    \item 
    \textit{Pointing and navigation will be difficult to impossible with the current ephemeris accuracy.} This was discussed in Subsection \ref{subsec:Ephemeris-errors}. The Proxima b ephemeris definitely needs work, but can be much improved before the first Proxima expedition is launched. It may be necessary to launch a precursor expedition one or more years before the first expedition targeted at Proxima b, specifically to improve the Proxima b orbital ephemeris. Note that this expedition would have to relay its information to successor expeditions without assistance from Earth, to avoid 8 year delays in updating the orbital ephemeris.
    \item 
    \textit{Dust and radiation can damage optics and electronics.} This can be reduced during the voyage by traveling edge-on, as discussed in Subsection \ref{subsec:damage-reduction}. Galactic cosmic rays (GCR) with energies $\gtrsim$ 20 MeV would not be significantly aberrated forward and thus would still strike all the probes. Although there are certainly spacecraft  functioning after 20 years of GCR exposure, this hazard requires more work before the first expeditions are launched.  
    \item 
    \textit{Data return will be difficult across interstellar distances.} This is discussed in Section \ref{sec:mission-to-Proxima} and will also require the data selection tools discussed in Section \ref{Sec:data-selection}. Although the techniques of developing swarm coherence and agent-based data selection certainly require work, there seems to be no fundamental limitation to the return of gigabytes of data over interstellar distances with large swarms of laser-sail spacecraft. 
\end{itemize}

\section{Conclusions}

Gram-scale interstellar probes pushed by laser light are likely to be the only technology capable of reaching another star this century; once
the laser infrastructure to do this is developed large 
expeditions of as many as a thousand probes could be sent
to 
any nearby nomadic planets (which should be present closer than Proxima and would make natural initial targets for precursor  missions) followed by expeditions to 
interstellar targets such as the Proxima Centauri system.
Both the transverse and the radial ephemeris errors for Proxima b are currently too large to realistically expect the swarm to come close to Proxima~b;  astrometry of the Proxima   system will have to be improved before these missions are sent.

Exploration
by coherent swarms of large numbers of small probes
offers numerous advantages \citep{Faria-et-al-2022-a,Eubanks-et-al-2023-a,Eubanks-et-al-2024-b}; 
a wide distribution of
probes provides redundancy, multiple views of the target,
and  a high probability that some probes will
pass close to the target planet.
As interstellar swarms will collect orders of magnitude more data than can be returned, and as the selection of data to return must be autonomous, the data selection system is a critical mission element. Agentic data selection systems should be developed to tailor this data selection to the needs of the mission science traceability matrix.

Picospacecraft swarms targeted at Proxima b will enable close passages to the planet, and thus high resolution gigapixel imaging of the target, substantially better that the kilopixel imaging that should eventually be possible with extremely large terrestrial optical telescope, or the megapixel imaging enabled by the solar gravitational lens \citep{Turyshev-2026-a}. 

A near-relativistic swarm mission could provide a strong initial survey of Proxima b as an exoplanet and should be able to detect Proxima b biosignatures and technosignatures, should these be present.  With this technology it should also be possible to perform in this century \textit{in situ} exploration by swarms of small spacecraft of  the entire Alpha Centauri system together with the Barnard's Star, Wolf 359  and Lalande 21185 systems, some brown dwarf systems such Luhman 16,  and any nearby nomadic planets.

\section{Acknowledgments}
This work was supported in part by  
the NASA Innovative Advanced Concepts (NIAC) contract 
80NSSC24K0649 ``Swarming Proxima Centauri: Coherent Picospacecraft Swarms Over Interstellar Distances,'' and also in part by a Breakthrough Starshot Foundation award to team \# 15 of the Breakthrough Starshot Communications Group.  Part of the work described here was also carried out at the Jet Propulsion Laboratory, California Institute of Technology, Pasadena, California, under a contract with the National Aeronautics and Space Administration.

\bibliographystyle{aasjournal}
\bibliography{Breakthrough,BTSScience2023}

\end{document}